\documentclass[12pt]{article}

\def\be{\begin{equation}}
\def\ee{\end{equation}}
\def\bea{\begin{eqnarray}}
\def\eea{\end{eqnarray}}

\def\e{{\rm e}}
\def\Tr{{\rm Tr}}
\def\lra{\longrightarrow}
\def\dr{{\rm d}^3\!\!\;r}
\def\d{{\rm d}}
\def\rhat{{\mathbf{\hat{r}}}}
\def\ehat{{\mathbf{\hat{e}}}}

\usepackage{graphicx}
\usepackage{bm}
\usepackage{amsmath}
\usepackage{amssymb}

\begin{document}

\begin{flushright}
hep-th/0503078
\end{flushright}

\pagestyle{plain}

\begin{center}
\vspace{2.5cm} {\Large {\bf On The Bound States Of Photons In Noncommutative Quantum Electrodynamics}}

\vspace{1cm}

Amir H. Fatollahi$^{~(1}$  \hspace{3mm} and \hspace{3mm} Abolfazl Jafari$^{~(2}$

\vspace{.5cm}

{\it 1) Mathematical Physics Group, Department of Physics, Alzahra University, \\ P. O. Box 19938, Tehran 91167, Iran}

\vspace{.3cm}

{\it 2) Institute for Advanced Studies in Basic Sciences (IASBS),\\
P. O. Box 45195, Zanjan 1159, Iran}

\vspace{.3cm}

\texttt{fatho@mail.cern.ch\\
jabolfazl@iasbs.ac.ir}

\vskip .5 cm
\end{center}

\begin{abstract}
We consider the possibility that photons of noncommutative QED can make bound states. Using the
potential model, developed based on the constituent gluon picture of QCD glue-balls,
arguments are presented in favor of existence of these bound states. The basic ingredient of potential
model is that the self-interacting massless gauge particles may get mass by inclusion non-perturbative
effects.
\end{abstract}

\vspace{2cm}


\newpage

\section{Introduction}
In Abelian gauge theories on ordinary space-time, there is no
self-coupling between the gauge fields. The best known
example is the quantum theory of interaction between electric
charges and photons, the so-called QED. The situation is different in
non-Abelian theories, and due to the commutator term in
the field strength, these theories are involved by direct
interaction between the quanta of gauge fields. It is now widely
believed that the strong interaction is described by a non-Abelian gauge
theory accompanied with proper matter fields as quarks, the so-called QCD.
As gluons are the quanta of QCD gauge field, from the very beginning this
possibility was considered that if gluons can make bound states free from valance quarks, the so-called glue-balls.
Although the properties of glue-balls have been studied for a long time,
their existence have not still been approved experimentally.

Recently a great interest has been appeared to study field theories on spaces
whose coordinates do not commute. These spaces, as well as the field theories
defined on them, are known under the names of noncommutative spaces and theories.
In contrast to QED on ordinary space-time, as we briefly review in next section,
noncommutative QED is involved by direct interactions between photons.
Interestingly one finds the situation very reminiscent to
that of non-Abelian gauge theories, and then the question is whether there
are some kinds of bound states in analogy with glue-balls of QCD, here might
be called ``photo-ball"s. It is this question that we consider in this work.
Our approach to study photo-balls is based on one of methods that has been
developed for glue-balls. As glue-balls are non-perturbative in nature,
there is still no systematic way for calculation of their properties from
the first principles of QCD. Instead, among the years many approaches are
developed for extracting glue-ball's properties, though each approach is
based on expectations or estimate calculations.

Among many others, one approach for studying properties of
glue-balls has been the so-called constituent gluon model.
In any study of bound state of gluons, one is encountered
with a situation in which gluons, though at first were introduced
massless to the Lagrangian, are bound and do not disjoint
to propagate to infinity. Correspondingly, it is argued that quantum fluctuations
around a charged particle, that should be treated non-perturbatively in QCD,
can make an accompanied cloud for it, causing a dynamically generated mass \cite{corn1,corn2}.
Accordingly, it appears very useful to define constituent
quarks and gluons, for which we assume a mass of order of
bound states of the theory, while their Lagrangian counterparts
may be massless or almost massless. As extracting the masses of
constituent particles from first principles is not yet done in
a satisfactory way, the best evaluations come from estimations
based on general considerations, phenomenology or lattice calculations.

Once one accepts that a glue-ball is a bound state of constituent
gluons, the question is about the effective theory that captures
the interaction between them. One approach is to consider
constituent gluons as massive quanta of an effective gauge theory.
It needs some kinds of proof, but hopefully this effective gauge theory
has the same qualitative features of the true (massless) theory, but in non-perturbative
regime \cite{corn1,corn2}. Since it is believed that the main contribution to the mass of a
glue-ball is coming from the constituent masses of gluons, it is
expected that constituent gluons move non-relativistically inside
the glue-ball, and so perturbative calculations for finding the
effective potential should be done in non-relativistic regime.
Having the effective potential at hand, by studying the
Schrodinger-type equations, one can have estimations about mass or
size of glue-balls. It is the heart of the potential model
approach for studying the properties of the glue-balls \cite{corn-soni,hou-luo-wong,hou-wong}.

There are two related issues when we are considering the
effective gauge theory of constituent gluons as massive vector
particles. First, it is known that the gauge symmetry is lost via
the mass term, and the second, massive gauge theories are known
to be perturbatively non-renormalizable. Here we give comments on
these issues \cite{corn1,corn2}. The non-renormalizability of massive
gauge theories is under this assumption that the mass in the theory appears
as a fixed parameter, surviving at large momentum. In fact the
insufficient decrease of propagator of a massive vector particle
at large momentum, due to simple power counting, suggests that
the theory can not be renormalizable. But the situation might be
different in a theory with constituent mass. At very large
momentum, where coupling constant is small due to asymptotic
freedom, the perturbation is valid and gluons appear as massless
particles. So the mass of constituent gluon, which is generated
dynamically, depends on momentum and vanishes at large momentum.
In a theory for gluons, it is argued that if one can keep the
dependence of constituent mass on momentum, which of course is
possible only by including the non-perturbative effects, the
theory may appear to be non-perturbatively renormalizable.

Although the argument above is for a model involved by dynamically generated mass,
due to lack of a systematic treatment of non-perturbative effects, much
can be learned via a kinematical description of gluon mass \cite{corn2};
it is to assume mass as a fix parameter, though the problem still remains with
local gauge symmetry. To overcome this problem, there is a prescription that we
review briefly in below. Let us consider firstly the case
with Abelian symmetry, and QCD case appears just as a generalization \cite{corn1,corn2}.
The starting point is the theory given by Lagrangian density:
\bea
L=-\frac{1}{4} F_{\mu\nu} F^{\mu\nu} +\frac{1}{2} m^2 \Big(A_\mu-\frac{1}{g}\partial_\mu\varphi\Big)^2
\eea
in which $g$ is the coupling constant, and $\varphi$ is a scalar field. The Lagrangian
is invariant under local symmetry
\bea
A'_\mu=A_\mu-\frac{1}{g}\partial_\mu \Lambda,\;\;\; \varphi'=\varphi-\Lambda,
\eea
with $\Lambda$ as symmetry transformation parameter. Now we
see although the gauge field has got mass, the local symmetry
is kept. Of course we mention that giving mass is done with
the price of introducing an extra scalar field. We have another
example of this observation in spontaneous symmetry breaking mechanism,
in which we are left with Goldstone bosons. In fact, this extra
scalar field, {\it just} like their Goldstone boson counterparts,
does not appear in the $S$-matrix, {\it i.e.} as external legs of diagrams.
One may insert the scalar field into the Lagrangian via
the obvious solution:
\bea
\varphi=g\frac{1}{\Box}\partial\cdot A,
\eea
getting a non-local but still gauge invariant theory involving only $A_\mu$.
It is the generalization of this mechanism that has been used for the case of
constituent gluon description of QCD glue-balls \cite{corn-soni,hou-luo-wong,hou-wong}, and here
we use for photo-balls of noncommutative QED. The starting point for
the QCD case is the Lagrangian density \cite{corn1,corn2}:
\bea
L=-\frac{1}{4}\Tr\big( F^{\mu\nu}F_{\mu\nu}\big)
+\frac{1}{2} m^2 \Tr
\Big(A_\mu-\frac{1}{g}V(\varphi)\partial_\mu V^\dagger(\varphi)\Big)^2
\eea
in which
\bea
V(\varphi)=\exp\Big[\frac{i}{2}\sum_a T^a\varphi_a\Big]
\eea
and $F_{\mu\nu}=\partial_{\mu}A_{\nu}-\partial_{\nu}A_{\mu}-ig[A_{\mu},A_{\nu}]$,
with $[T^a,T^b]=if^{ab}_cT^c$. The action is invariant under
\bea
A'_\mu &=&UA_\mu U^{-1}+\frac{1}{g}U\partial_\mu U^{-1},\nonumber\\
V(\varphi')&=&U V(\varphi),
\eea
in which $U=U(x)$ is unitary matrix defining the transformation.
Again one can insert the scalar fields into the Lagrangian via power
series solution in $g$ \cite{corn1,corn2}:
\bea
\varphi_a=g\frac{1}{\Box}\partial\cdot A_a - g^2 [\cdots]_a .
\eea
As mentioned above the extra scalars do not appear as external legs of diagrams,
but the situation is even simpler as far as one considers just the tree diagrams,
in which one can ignore the scalars. So for tree diagrams, and in proper gauge,
the Lagrangian density in use is practically \cite{corn2,hou-luo-wong,hou-wong}:
\bea
L=-\frac{1}{4}F^{a\mu\nu}F^a_{\mu\nu}+\frac{1}{2} m^2 A^{a\mu}A^a_\mu,
\eea
simply as a gauge theory for massive gluons.

In the non-relativistic limit the potential can be read off from
the total invariant amplitude ${\cal M}_{fi}$ via the Fourier transform
\cite{corn-soni,hou-luo-wong,hou-wong}:
\bea\label{potential}
V(\mathbf{r})=\int
\frac{{\rm d}^{3}q}{8\pi^{3}}\frac{i\e^{i\mathbf{q}\cdot\mathbf{r}}}{4\sqrt{E_{1}E_{2}E_{3}E_{4}}}\,i{\cal M}_{fi}
\eea
in which $\mathbf{q}$ is the momentum transferred between the in-coming particles.
The total invariant amplitude gets contribution at tree level
from the s-, t-, u- channels, and the so-called seagull (s.g.)
diagram, coming from 4-gluon vertex of QCD \cite{corn-soni,hou-luo-wong,hou-wong}. In non-relativistic
limit it can be shown that the s-channel's contribution is ignorable,
to get the final expression \cite{corn-soni,hou-luo-wong,hou-wong}:
\bea\label{t4gluons}
i{\cal M}_{fi}&=&\frac{ig^{2}f^{ace}f^{bde}}{\mathbf{q}^2+m^{2}}\times
\Big(4m^{2}+3\mathbf{q}^{2}-2\mathbf{S}^{2}\mathbf{q}^{2}
+2(\mathbf{S\cdot q})^{2}
+6i\mathbf{S}\cdot(\mathbf{q}\times\mathbf{p}_{i})\Big)
\nonumber\\
&~&-ig^{2}\Big(f^{abe}f^{cde}-f^{ace}f^{bde}\big(\frac{1}{2}\mathbf{S}^{2}-2\big)\Big).
\eea
By above the gluon-gluon potential in one-gluon-exchange (oge) approximation
appears in the form \cite{corn-soni,hou-luo-wong,hou-wong}:
\bea\label{oge}
V_{\rm oge}(\mathbf{r})\!\!\!\!&=&\!\!\!\!\frac{-g^{2}f^{ace}f^{bde}}{4\pi}\bigg\{\Big[\frac{1}{4}+\frac{1}{3}
\mathbf{S}^{2}-\frac{3}{2m^{2}}(\mathbf{L}\cdot\mathbf{S})\nabla_{r}\bigg.
-\frac{1}{2}m^{2}\big((\mathbf{S}\cdot\bm{\nabla})^{2}
-\frac{1}{3}\mathbf{S}^{2}\nabla^2\big)\Big]\frac{\e^{-mr}}{r}
\nonumber\\
&&-\big(1-\frac{5}{6}\mathbf{S}^{2}\big)\frac{\pi}{m^2}\delta^{3}(\mathbf{r})\bigg\}
+\frac{g^{2}f^{abe}f^{cde}}{4\pi}\frac{\pi}{m^2}\delta^{3}(\mathbf{r})
\eea
in which $\nabla_r=r^{-1}\partial_r$.

The organization of the rest of this work is as follows. In Sec. 2 we review some basic features of
canonical noncommutative spaces, and also the field theories defined on them, specially noncommutative QED.
We also remark on some aspects of noncommutative QED that make this theory in some extents
similar to QCD. Sec. 3 is devoted to extracting the effective potential between massive photons, by studying
the non-relativistic behavior of their scattering. Sec. 4 mainly contains the dynamics of photons under
the effective potential obtained in Sec. 3. The existence proof of bound states is also presented in Sec. 4.
Sec. 5 is for conclusion and discussions.

\section{Noncommutative Space-Time And QED}
In last years a great attention is appeared in formulation and
studying field theories on noncommutative spaces. One of the
original motivations has been to get "finite" field theories via
the intrinsic regularizations which are encoded in some of noncommutative
spaces \cite{snyder}. The other motivation comes back to the natural
appearance of noncommutative spaces in some areas of physics, and the recent
one in string theory. It has been understood that string theory is
involved by some kinds of noncommutativities; two examples are,
1) the coordinates of bound-states of $N$ D-branes are presented
by $N\times N$ Hermitian matrices \cite{9510135}, and 2) the
longitudinal directions of D-branes in the presence of B-field
background appear to be noncommutative, as are seen by the ends of open
strings \cite{9908142,cds,jab1}. In the latter case for constant
background one simply gets the canonical noncommutative
space-time, introduced by commutation relations for coordinates
as:
\bea
[\hat{x}^{\mu},\hat{x}^{\nu}]=i\theta^{\mu\nu}=\frac{i}{\omega^{2}}c^{\mu\nu}
\eea
in which the parameter $\omega$ has the dimension of
energy, and signifies the scale where noncommutative effects become
relevant. The $c^{\mu\nu}$ is a real antisymmetric matrix with
elements of order one. Now since the coordinates do not commute,
any definition of functions or fields should be performed under a
prescription for ordering of coordinates, and one choice can be
the symmetric one, the so-called Weyl ordering. To any function
$f(x)$ on ordinary space, one can assign an operator $\hat{O}_f$
by
\bea
\hat{O}_f\big(\hat{x}\big):=\frac{1}{(2\pi)^n} \int
{\rm d}^nk \; \tilde{f}(k)\; \e^{-ik\cdot \hat{x}}
\eea
in which $\tilde{f}(k)$ is the Fourier transform of $f(x)$ defined by
\bea
\tilde{f}(k)=\int {\rm d}^nx \; f(x)\; \e^{ik\cdot x}.
\eea
Due to presence of the phase $\e^{-ik\cdot \hat{x}}$ in definition of
$\hat{O}_f$, we recover the Weyl prescription for the coordinates.
In a reverse way we also can assign to any symmetrized operator a
function or field living on the noncommutative plane. Also, we can
assign to product of any two operators $\hat{O}_f$ and $\hat{O}_g$
another operator as
\bea
\hat{O}_f\cdot \hat{O}_g =:\hat{O}_{f\star g}
\eea
in which $f$ and $g$ are multiplied under the so-called $\star$-product defined by
\bea
(f\star g)(x)=\exp\big(\frac{i\theta^{\mu\nu}}{2}\partial_{x_\mu}\partial_{y_\nu}\big)f(x)g(y)\mid_{y=x}
\eea
By these all one learns how to define physical theories on noncommutative space-time, and
eventually it appears that the noncommutative field theories are defined by actions that
are essentially the same as in ordinary space-time, with the exception that the products
between fields are replaced by $\star$-product; see \cite{reviewnc} as review. Though $\star$-product itself is
not commutative ({\it i.e.}, $f \star g \neq g \star f$) the following identities make some of calculations
easier:
\bea
&~&\int f\star g\;{\rm d}^{n}x=\int g\star f\; {\rm d}^{n}x=\int fg\;{\rm d}^{n}x\nonumber\\
&~&\int f\star g\star h\;{\rm d}^{n}x=\int f (g\star h)\;{\rm d}^{n}x=\int (f\star g) h\;{\rm d}^{n}x\nonumber\\
&~&\int f\star g\star h\;{\rm d}^{n}x=\int h\star f\star g \;{\rm d}^{n}x=\int g\star h\star f\;{\rm d}^{n}x\nonumber
\eea
By the first two ones we see that, in integrands always one of the stars can be removed.
Besides it can be seen that the $\star$-product is associative, {\it i.e.},
$f\star g\star h=(f\star g)\star h= f\star (g\star h)$, and so it is not important which
two ones should be multiplied firstly.

The pure gauge field sector of noncommutative QED is defined by the action
\bea
S_{\rm gauge-field}&=&-\frac{1}{4}\int {\rm d}^{4}x \;F_{\mu\nu}\star F^{\mu\nu}
\nonumber\\
&=&-\frac{1}{4}\int {\rm d}^{4}x \;F_{\mu\nu}F^{\mu\nu}
\eea
in which the field strength $F_{\mu\nu}$ is
\bea\label{field}
F_{\mu\nu}=\partial_{\mu}A_{\nu}(x)-\partial_{\nu}A_{\mu}(x)-ie[A_{\mu}(x),A_{\nu}(x)]_{\star}
\eea
by definition $[f,g]_{\star}=f\star g-g\star f$.
We mention $[x^\mu,x^\nu]_{\star}=i\theta^{\mu\nu}$. The action above is invariant under local gauge symmetry transformations
\bea\label{trans}
A'_{\mu}(x)\!=\! U\star A_{\mu}(x)\star U^{-1}+\frac{i}{e}U\star
\partial _{\mu}U^{-1}
\eea
in which $U=U(x)$ is the $\star$-phase, defined by a function $\lambda(x)$
via the $\star$-exponential:
\bea\label{starphase}
&~&U(x)=\exp_{\star}(i\lambda)=1+i\lambda-\frac{1}{2}\lambda\star\lambda+\cdots,\\
&~&U\star U^{-1}=U^{-1}\star U=1
\eea
in which $U^{-1}=\exp_{\star}(-i\lambda)$. Under above transformation, the field strength transforms as
\bea
F_{\mu\nu}&\lra& F^{\prime}_{\mu\nu}=U\star F_{\mu\nu}\star U^{-1}
\eea
We mention that the transformations of gauge field as well as the field strength
look like to those of non-Abelian gauge theories. Besides we see that the
action contains terms which are responsible for interaction
between the gauge particles, again as the situation we have in non-Abelian gauge
theories. We see how the noncommutativity of coordinates induces properties on fields and
their transformations, as if they were belong to a non-Abelian theory;
the subject that how the characters of coordinates and fields may be related to each other
is discussed in \cite{fath}. These observations make it reasonable to study whether and how the photons can make
bound states in such a theory.

There is another observation that promotes the formal similarities of noncommutative and non-Abelian
theories to their behaviors, that is the negative sign of $\beta$-function, which manifests that
these theories are asymptotically free \cite{martin-ruiz, jab2}. By this it is more reasonable to see
if the techniques developed for QCD purposes can also be used for noncommutative QED.

The phenomenological implications of possible noncommutative coordinates have been the subject
of a very large number of research publications in last years. Among many others, here we can give just a
brief list of works, and specially those concerning the phenomenological implications of noncommutative QED.
The effect of noncommutativity of space-time is studied for possible modifications
that may appear in high energy scattering amplitudes of particles \cite{phen-nc1}, in energy levels of
light atoms \cite{phen-nc2, phen-nc3}, and anomalous magnetic moment of electron \cite{jab3}.
The ultra-high energy scattering of massless photons of noncommutative QED is considered
in \cite{mahajan} and the tiny change in the total amplitude is obtained as a function of the total energy.
Some other interesting features of noncommutative ED and QED are discussed in \cite{morencqed}.
The issue of formation of new bound states by space-time noncommutativity has been considered in \cite{yurov}.

\section{Massive Noncommutative QED And Effective \\ Potential Between Photons}
\subsection{Massive Photon-Photon Scattering Amplitude}
Now here, following the procedure developed for QCD case, we give mass to photons of
noncommutative QED. As described this is done by introducing an extra
scalar field, getting the Lagrangian density:
\bea
L=L_{\rm gauge-field}+\frac{m^{2}}{2}\Big(A_{\mu}+\frac{i}{e}\partial
_{\mu}V(\varphi)\star V^{-1}(\varphi)\Big)_\star^2
\eea
in which $(\cdots)_\star^2=(\cdots)\star(\cdots)$, and $V(\varphi)$
is the $\star$-phase defined by the scalar field $\varphi$; see (\ref{starphase}).
The action defined by above Lagrangian is invariant under transformations
\bea
A'_{\mu}(x)\!&=&\! U\star A_{\mu}(x)\star U^{-1}+\frac{i}{e}U\star\partial _{\mu}U^{-1}\nonumber\\
V(\varphi')\!&=&\!U\star V(\varphi)
\eea
in which $U$ is the same in (\ref{trans}). Now we just list the Feynman rules \cite{corn2}\cite{jab2,jab3}. We choose
the gauge in which the propagator takes the form:
\bea
iD^{\mu\nu}\big(p)=\frac{-ig^{\mu\nu}}{p^{2}-m^{2}}
+\frac{ip^{\mu}p^{\nu}}{(p^{2}-m^{2})m^{2}}.
\eea
In the non-relativistic limit we have for momentum and polarization vectors \cite{hou-luo-wong,hou-wong}
\bea\label{momnonrev}
&&p^{\mu}=\big(m+\frac{\mathbf{p^{2}}}{2m},\mathbf{p}\big)\\
\label{polnonrev}
&&\epsilon ^{\mu}=\big(\frac{\mathbf{p}\cdot
\mathbf{e}}{m},\mathbf{e}+\frac{\mathbf{p}\cdot\mathbf{e}}{2m^{2}}\mathbf{p}\big)
\eea
in which $\textbf{e}$ is a 3-vector satisfying $\mathbf{e}^* \cdot \mathbf{e}=1$. From Lorentz
condition \cite{corn-soni,hou-luo-wong,hou-wong}, we have
\bea
p\cdot \epsilon=p^{\mu} \epsilon_{\mu}=0
\eea
In this work we assume for the signature of metric $g_{\mu\nu}=(+1,-1,-1,-1)$.
\begin{figure}[tbp]
\begin{center}
\includegraphics[width=0.5\columnwidth]{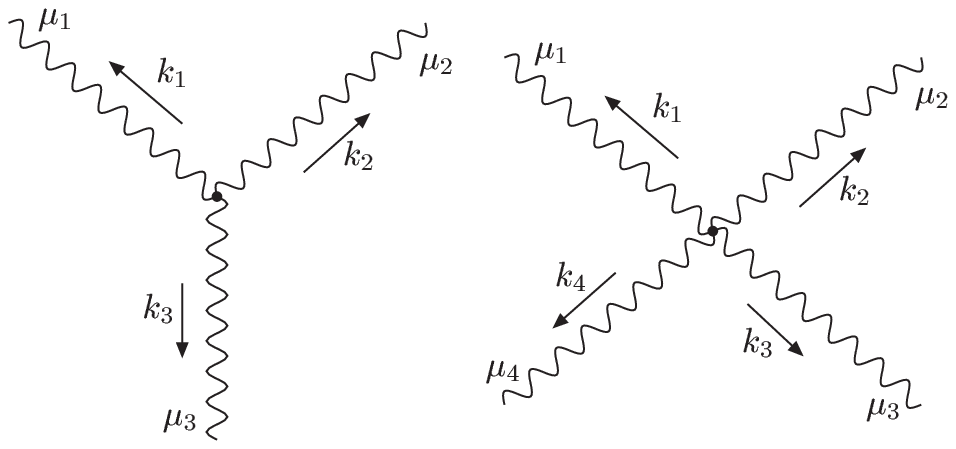}
\caption{3- and 4- photon vertices of noncommutative QED.}
\label{fig1}
\end{center}
\end{figure}
As in this work we restrict ourselves to tree diagrams, after removing
one $\star$, and by Lorentz condition, we practically are using the Lagrangian \cite{corn2}
\bea
L=-\frac{1}{4}F^{\mu\nu}F_{\mu\nu}+\frac{1}{2} m^2 A^{\mu}A_{\mu},
\eea
with $F_{\mu\nu}$ defined in (\ref{field}).
There are 3- and 4- photon vertices given in Fig. 1.
As in this work we consider noncommutativity just in spatial directions, that is
assuming $\theta^{0i}=0$ for $i=1,2,3$, we have for the vertex-functions \cite{jab2,jab3}:
\bea
\Gamma^{\mu_1\mu_2\mu_3}_{k_1,k_2,k_3}=
-2e\sin\!\big(\frac{\mathbf{k}_{1}\ltimes\mathbf{k}_{2}}{2}\big)
\Big[(k_{1}-k_{2})^{\mu_{3}}g^{\mu_{1}\mu_{2}}
+(k_{2}-k_{3})^{\mu_{1}}g^{\mu_{2}\mu_{3}}+ (k_{3}-k_{1})^{\mu_{2}}g^{\mu_{3}\mu_{1}}\Big]
\eea
and
\bea
\Gamma^{\mu_1\mu_2\mu_3\mu_4}_{k_1,k_2,k_3,k_4}=-4ie^{2}\!\!\!\!\!\!\!\!\!
&&\Big[\sin\!\big(\frac{\mathbf{k}_{1}\ltimes \mathbf{k}_{2}}{2}\big)\sin\!\big(\frac{\mathbf{k}_{3}\ltimes
\mathbf{k}_{4}}{2}\big)\big(g^{\mu_{1}\mu_{3}}g^{\mu_{2}\mu_{4}}\!\!-\!g^{\mu_{1}\mu_{4}}g^{\mu_{2}\mu_{3}}\!\big)
\nonumber\\
&&+\sin\!\big(\frac{\mathbf{k}_{3}\ltimes \mathbf{k}_{1}}{2}\big)\sin\!\big(\frac{\mathbf{k}_{2}\ltimes
\mathbf{k}_{4}}{2}\big)\big(g^{\mu_{1}\mu_{4}}g^{\mu_{2}\mu_{3}}\!\!-\!g^{\mu_{1}\mu_{2}}g^{\mu_{3}\mu_{4}}\!\big)
\nonumber\\
&&+\sin\!\big(\frac{\mathbf{k}_{1}\ltimes \mathbf{k}_{4}}{2}\big)\sin\!\big(\frac{\mathbf{k}_{2}\ltimes
\mathbf{k}_{3}}{2}\big)\big(g^{\mu_{1}\mu_{2}}g^{\mu_{3}\mu_{4}}\!\!-\!g^{\mu_{1}\mu_{3}}g^{\mu_{2}\mu_{4}}\!\big)\Big]
\eea
in which $\mathbf{a}\ltimes\mathbf{b}\equiv \theta^{ij} a_i b_j$, and
the momenta and indices are given in Fig. 1. Also in each vertex energy-momentum conservation
should be understood.

Although there are four diagrams at tree level, those coming from s-, t-, u- channels, and seagull diagram (Fig. 2),
when extracting the potential, by the properly symmetrized wave function for identical particle
systems, the ``exchange" or ``symmetry" diagrams are automatically taken care of, causing that only one of t- and u- channels'
contributions should be added to others' contributions \cite{hou-luo-wong}. For s-channel we have
\bea
i{\cal M}^{\rm s}_{fi}&=&4e^{2}\sin\!\big(\frac{\mathbf{p_{1}}\ltimes\mathbf{k}}{2}\big)
\sin\!\big(\frac{\mathbf{p_{3}}\ltimes\mathbf{k}}{2}\big)\epsilon_{1\mu_1}\epsilon_{2\mu_2}  \Gamma^{\mu_1\mu_2\mu}_{-p_1,-p_2,k}
\nonumber\\
&&\!\Big(\frac{-ig_{\mu\nu}}{k^{2}-m^{2}}   +\frac{ik_{\mu}k_{\nu}}{(k^{2}-m^{2})m^{2}}\Big)
\;\Gamma^{\nu\mu_3\mu_4}_{-k,p_3,p_4}\;\epsilon_{3\mu_3}^{*}\epsilon_{4\mu_4}^{*}
\eea
In s-channel we have $k=p_{1}+p_{2}=p_{3}+p_{4}$, and by using it we get
\bea
i{\cal M}^{\rm s}_{fi}&=&-4ie^{2}\frac{\sin\!\big(\frac{\mathbf{p_{1}}\ltimes
\mathbf{k}}{2}\big)\sin\!\big(\frac{\mathbf{p_{3}}\ltimes
\mathbf{k}}{2}\big)}{q^{2}-m^{2}}
\Big(\epsilon_{1}\cdot \epsilon_{2}(p_{1}-p_{2})_{\rho}+2\epsilon
_{2\rho}p_{2}\cdot\epsilon_{1}-2\epsilon_{1\rho}p_{1}\cdot\epsilon_{2}\Big)
\nonumber\\
&&\Big(\epsilon^{*}_{3}\cdot \epsilon ^{*}_{4}(p_{4}-p_{3})^{\rho}-2\epsilon^{*\rho}_{4}
p_{4}\cdot\epsilon^{*}_{3}+2\epsilon^{*\rho}_3 p_{3}\cdot\epsilon^{*}_{4}\Big)
\eea

\begin{figure}[tbp]
\begin{center}
\includegraphics[width=0.6\columnwidth]{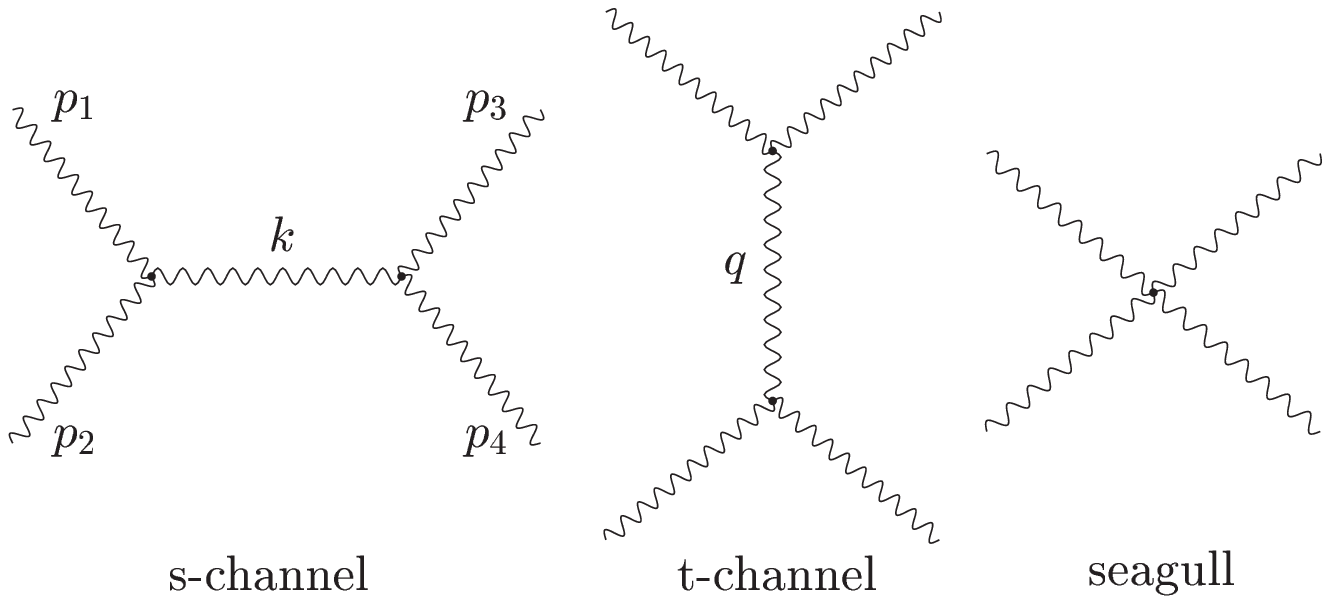}
\caption{s-channel, t-channel, and seagull diagrams.}
\label{fig2}
\end{center}
\end{figure}

From the Lorentz condition the last term in propagator is omitted.
By replacing the non-relativistic limit of $\epsilon$'s, we see that,
even without considering coefficient involving sin($\cdots$), the leading order contribution of
s-channel is order of $|\mathbf{p}|^{2}\ll m^2$, that we ignore comparing the
zeroth orders. This observation is exactly the same that happens in the QCD case
\cite{corn-soni,hou-luo-wong,hou-wong}.

Now we come to t-channel,
\bea
i{\cal M}^{\rm
t}_{fi}\!=\!\!-4e^{2}\!\sin\!\big(\frac{\mathbf{p_{1}}\ltimes
\mathbf{q}}{2}\big)\!\sin\!\big(\frac{\mathbf{p_{2}}\ltimes
\mathbf{q}}{2}\big)
\!\Big[g_{\mu\lambda}(p_{1}+p_{3})_{\rho}+g_{\lambda\rho}(p_{1}-2p_{3})_{\mu}+
g_{\rho\mu}(p_{3}-2p_{1})_{\lambda}\Big]
\nonumber\\
\epsilon^{\mu}_{1}\epsilon^{*\lambda}_{3}\frac{-i\big(g^{\rho\delta}
-\frac{q^{\rho}q^{\delta}}{m^{2}}\big)}{q^{2}-m^{2}}\epsilon^{\nu}_{2}\epsilon^{*\sigma}_{4}
\Big[g_{\nu\sigma}
(p_{2}+p_{4})_{\delta}+g_{\sigma\delta}(p_{2}-2p_{4})_{\nu}+g_{\delta\nu}(p_{4}-2p_{2})_{\sigma}\Big]
\eea
in which $q=p_{3}-p_{1}=p_{2}-p_{4}$. It appears to be useful if we define \cite{hou-luo-wong,hou-wong}
\bea
\langle 3|J_{\rho}|1\rangle =\epsilon ^{*}_{3}\cdot \epsilon
_{1}(p_{1}+p_{3})_{\rho}-2\epsilon ^{*}_{3\rho}p_{3}\cdot
\epsilon _{1}-2\epsilon_{1\rho}p_{1}\cdot \epsilon^{*}_{3}
\eea
and a similar one for $\langle 4|J^\delta|2\rangle$, and
since $\langle 3|J_{\rho}|1\rangle q^{\rho}=\langle 4|J_{\delta}|2\rangle q^{\delta}=0$, thus,
\bea
i{\cal M}^{\rm t}_{fi}\!=\!4ie^{2}\frac{\sin\!\big(\frac{\mathbf{p_{1}}\ltimes
\mathbf{q}}{2}\big)\sin\!\big(\frac{\mathbf{p_{2}}\ltimes
\mathbf{q}}{2}\big)}{q^{2}-m^{2}}\langle 3|J_{\rho}|1\rangle \langle
4|J^{\rho}|2\rangle
\eea
We continue in the center-of-mass frame, for which
\bea
&&\mathbf{p}_{1}=-\mathbf{p}_{2}=\mathbf{p}_{i},\nonumber\\
&&\mathbf{p}_{3}=-\mathbf{p}_{4}=\mathbf{p}_{i}+\mathbf{q}=\mathbf{p}_{f},\nonumber\\
&&\mathbf{q}=\mathbf{p}_{3}-\mathbf{p}_{1}=\mathbf{p}_{f}-\mathbf{p}_{i}.
\eea
In the final expression we keep just the zeroth order of $|\mathbf{p}|$, although we
should be careful about $\mathbf{q}$ dependence, on which we integrate over to find
the large-distance behavior of the effective potential. So in the following steps we still should keep orders of
$|\mathbf{p}|^{2}$ in $\langle 3|J_{\rho}|1\rangle \langle 4|J^{\rho}|2\rangle$.

The next steps of calculations are essentially those done for glue-balls
\cite{corn-soni,hou-luo-wong,hou-wong}, that we present in below. Photon's spin is one, and so it is useful
to define the operators $\mathbf{S}=(S^{1},S^{2},S^{3}\big)$ as
\bea
S^{1}=\left(
\begin{array}{ccc}
  0 & 0 & 0 \\
  0 & 0 & -i \\
  0 & i & 0 \\
\end{array}
\right),\;\;\;
S^{2}=\left(
\begin{array}{ccc}
  0 & 0 & i \\
  0 & 0 & 0 \\
 -i & 0 & 0 \\
\end{array}
\right),
\;\;\;S^{3}=\left(
\begin{array}{ccc}
  0 & -i & 0 \\
  i & 0 & 0\\
  0 & 0 & 0 \\
\end{array}
\right),
\eea
for which we have $[S^{i},S^{j}]=i\epsilon ^{ijk}S^{k}$.

Now we calculate the components of $\langle 3|J^\rho|1\rangle$.
Using (\ref{momnonrev}) and (\ref{polnonrev}) we get
\bea
\langle 3|J^{0}|1\rangle&=&-\big(2m+\frac{\mathbf{p}^{2}_{1}+\mathbf{p}^{2}_{3}}{2m}\big)
\big(\mathbf{e}^{*}_{3}\cdot \mathbf{e}_{1}\big)
\nonumber\\
&&+\frac{1}{m}   [-2(\mathbf{p}_{1}\cdot \mathbf{e}_{1})(\mathbf{p}_{3}\cdot
\mathbf{e}^{*}_{3})+(\mathbf{p}_{3}\cdot
\mathbf{e}_{1})(\mathbf{p}_{3}\cdot \mathbf{e}^{*}_{3})
+(\mathbf{p}_{1}\cdot
\mathbf{e}_{1})(\mathbf{p}_{1}\cdot \mathbf{e}^{*}_{3})]
\eea
For any two vectors $\mathbf{A}$ and $\mathbf{B}$ we mention
\bea
A_{i}B_{j}=\big[\mathbf{A\cdot B} 1\!\!1-\mathbf{(S\cdot B)(S\cdot A)}\big]_{ij}
\eea
(in the following we drop the identity operator $1\!\!1$). Using above we can write
\bea
\langle 3|J^{0}|1\rangle&=&-e^\dag
_{3}\big(2m+\frac{\mathbf{p}^{2}_{1}+\mathbf{p}^{2}_{3}}{2m}\big)e_{1}
\nonumber\\
&&+\frac{1}{m}e^\dag
_{3}\big[\mathbf{q}^{2}-(\mathbf{S} \cdot \mathbf{q})^{2}+(\mathbf{S}
\cdot \mathbf{p}_{1})(\mathbf{S}\cdot \mathbf{p}_{3})
-(\mathbf{S}\cdot \mathbf{p}_{3})(\mathbf{S}\cdot \mathbf{p}_{1})\big]e_{1}
\eea
in which $e_1$ and $e_3$ are column-matrix representation of 3-vectors $\mathbf{e}_1$ and $\mathbf{e}_3$ . Again using the identity
\bea
(\mathbf{S}\cdot \mathbf{p}_{1})(\mathbf{S} \cdot
\mathbf{p}_{3})-(\mathbf{S}\cdot \mathbf{p}_{3})(\mathbf{S} \cdot
\mathbf{p}_{1})
=p^{i}_{1}p^{j}_{3}[S^{i},S^{j}]
=i(\mathbf{p}_{1}\times
\mathbf{p}_{3})\cdot \mathbf{S}
\eea
we finally reach to
\bea
\langle 3|J^{0}|1\rangle\!\!&=&\!\!-e^\dag_{3}\Big[\big(-2m-\frac{\mathbf{p}^{2}_{1}+\mathbf{p}^{2}_{3}}{2m}\big)
+\frac{1}{m}\big[\mathbf{q}^{2}-(\mathbf{S}_{1}\cdot \mathbf{q})^{2}
+i(\mathbf{p}_{1}\times \mathbf{p}_{3})\cdot \mathbf{S}_{1}\big]\Big]e_{1}
\eea
in which $\mathbf{S}_{1}$ as first photon's spin-operator. Similarly,
\bea
\langle 4|J^{0}|2\rangle=-e^\dag_{4}\Big[\big(-2m-\frac{\mathbf{p}^{2}_{2}+\mathbf{p}^{2}_{4}}{2m}\big)
+\frac{1}{m}\big[\mathbf{q}^{2}-(\mathbf{S}_{2}\cdot \mathbf{q})^{2}
+i(\mathbf{p}_{2}\times \mathbf{p}_{4})\cdot \mathbf{S}_{2}\big]\Big]e_{2}
\eea
Now using $p\cdot\epsilon=0$, we can write the last two terms of
$\langle 3|J^{k}|1\rangle$ in terms of $q$, and so we get
\bea
\!\!\!\!\!\!\langle 3|J^{k}|1\rangle\!\!&=&\!\!-e^\dag_{3}(\mathbf{p}_{1}+\mathbf{p}_{3})^{k}e_{1}+2e^{*k}_{3}\mathbf{q}
\cdot e_{1}-2e^{k}_{1}\mathbf{q}\cdot e^{*}_{3}
\nonumber\\
\!\!&=&\!\!-e^\dag_{3}(\mathbf{p}_{1}+\mathbf{p}_{3})^{k}e_{1}+2e^{* i}_{3}[\delta ^{ik}q^{j}-\delta ^{jk}q^{i}] e^{j}_{1}
\eea
Let us define the matrices $B_1, B_2, B_3$ by their elements,
\bea
B^{ij}_k\equiv \delta ^{ik}q^{j}-\delta^{jk}q^{i}
\eea
and so for example we see for $B_1$ that
\bea
B_1=\left(
\begin{array}{ccc}
  0 & q^{2} & q^{3} \\
  -q^{2} & 0 & 0\\
  -q^{3} & 0 & 0 \\
\end{array}
\right)
=-i\big[(\mathbf{S}_1\times \mathbf{q})\big]_1
\eea
yielding $\delta^{ik}q^{j}-\delta^{jk}q^{i}=-i\big[(\mathbf{S}_{1}\times \mathbf{q})\big]^{ij}_k$.
By using the above we get
\bea
\langle 3|\mathbf{J}|1\rangle=e^\dag_{3}[-(\mathbf{p}_{1}+\mathbf{p}_{3})-2i(\mathbf{S}_{1}\times\mathbf{q})]e_{1}
\\
\langle 4|\mathbf{J}|2\rangle=e^\dag _{4}[-(\mathbf{p}_{2}+\mathbf{p}_{4})+2i\big(\mathbf{S}_{2}\times\mathbf{q})]e_{2}
\eea
By these all we have
\bea
\langle 3|J^{\rho}|1\rangle \langle 4|J_{\rho}|2\rangle&=&e^\dag_{3}e^\dag_{4}
\big[4 m^{2}\!+\!\mathbf{p}^{2}_{1}+\mathbf{p}^{2}_{2}+\mathbf{p}^{2}_{3}+\mathbf{p}^{2}_{4}-4\mathbf{q}^{2}
\nonumber\\
&&-(\mathbf{p}_{1}+\mathbf{p}_{3})\cdot(\mathbf{p}_{2}+\mathbf{p}_{4})
+2(\mathbf{S}_{1}\cdot \mathbf{q})^{2}+2(\mathbf{S}_{2}\cdot \mathbf{q})^{2}
\nonumber\\
&&-4(\mathbf{S}_{1}\times\mathbf{q})\cdot(\mathbf{S}_{2} \times\mathbf{q})
+2i(\mathbf{p}_{3} \times \mathbf{p}_{1})\cdot\mathbf{S}_{1}
+2i(\mathbf{p}_{4} \times \mathbf{p}_{2})\cdot\mathbf{S}_{2}
\nonumber\\
&&+2i(\mathbf{p}_{1}+ \mathbf{p}_{3})\cdot
(\mathbf{S}_{2}\times \mathbf{q})
-2i(\mathbf{p}_{2}+\mathbf{p}_{4})\cdot (\mathbf{S}_{1}\times \mathbf{q})\big]e_{1}e_{2}
\eea
in which we should care the order in multiplying the objects in above.
By conservation of energy we have
$\mathbf{p}^{2}_{1}+\mathbf{p}^{2}_{2}=\mathbf{p}^{2}_{3}+\mathbf{p}^{2}_{4}$,
and so in the center-of-mass frame
$\mathbf{p}^{2}_{1}=\mathbf{p}^{2}_{2}=\mathbf{p}^{2}_{3}=\mathbf{p}^{2}_{4}$.
Also since $(\mathbf{p}_{1}-\mathbf{p}_{3})(\mathbf{p}_{1}+\mathbf{p}_{3})=0$
and so $(\mathbf{p}_{1}+\mathbf{p}_{1}-\mathbf{q})\cdot(\mathbf{p}_{1}-\mathbf{p}_{1}+\mathbf{q})=0$
we obtain $\mathbf{p}_{i}\cdot \mathbf{q}=\frac{-1}{2}\mathbf{q}^{2}=-\mathbf{p}_{f}\cdot \mathbf{q}$.
By using these all we have $(\mathbf{p}_{1}+\mathbf{p}_{3})\cdot
(\mathbf{p}_{2}+\mathbf{p}_{4})=-4\mathbf{p}^{2}_{i}+\mathbf{q}^{2}$,
and so
\bea
\mathbf{p}^{2}_{1}+\mathbf{p}^{2}_{2}+\mathbf{p}^{2}_{3}+\mathbf{p}^{2}_{4}-4\mathbf{q}^{2}
-(\mathbf{p}_{1}+\mathbf{p}_{3})\cdot(\mathbf{p}_{2}+\mathbf{p}_{4})
=8\mathbf{p}^{2}_{i}-5\mathbf{q}^{2}
\eea
that we ignore $\mathbf{p}_i^2$ in comparison with $m^2$. For the terms involving
spin-operators defining the total spin $\mathbf{S}=\mathbf{S}_{1}+\mathbf{S}_{2}$, we get
$\mathbf{S}_{1}\cdot \mathbf{S}_{2}=\frac{1}{2}\mathbf{S}^{2}-2$, and thus
\bea
(\mathbf{S}_{1}\cdot \mathbf{q})^{2}+(\mathbf{S}_{2}\cdot
\mathbf{q})^{2}-2(\mathbf{S}_{1}\times
\mathbf{q})\cdot(\mathbf{S}_{2}\times\mathbf{q})
=(4-\mathbf{S}^{2})\mathbf{q}^{2}+(\mathbf{S}\cdot
\mathbf{q})^{2}
\eea
By two other relations
\bea
&&i( \mathbf{p}_{3}\times \mathbf{p}_{1})\cdot
\mathbf{S}_{1}+i(\mathbf{p}_{4}\times \mathbf{p}_{2})\cdot
\mathbf{S}_{2}=i(\mathbf{q}\times \mathbf{p}_{i})\cdot
\mathbf{S}\\
&&i(\mathbf{p}_{1}+\mathbf{p}_{3})\cdot (\mathbf{S}_{2}\times
\mathbf{q})-i(\mathbf{p}_{2}+\mathbf{p}_{4})\cdot
(\mathbf{S}_{1}\times \mathbf{q})
=2i \mathbf{S}\cdot
(\mathbf{q}\times \mathbf{p}_{i})
\eea
and using $q^2=0-{\bf q}^2$ and $\mathbf{p}_1=-\mathbf{p}_2=\mathbf{p}$, we obtain
\bea
i{\cal M}^{\rm t}_{fi}=4ie^2\frac{\sin^2\!\big(\frac{\mathbf{p}\ltimes
\mathbf{q}}{2}\big)}{\mathbf{q}^{2}+m^{2}}   \big[4m^{2}+3\mathbf{q}^{2}
-2\mathbf{S}^{2}\mathbf{q}^{2}+2(\mathbf{S}\cdot
\mathbf{q})^{2}+6i\mathbf{S} \cdot (\mathbf{q} \times
\mathbf{p})\big]+O(\mathbf{p}^2)
\eea
We mention that the kinematical dependence of t-channel amplitude, given by terms in $[\cdots]$,
no surprisingly is exactly that for gluons, presented in relation (\ref{t4gluons}). In fact the only difference between the case
of gluons and photons in noncommutative QED is in the pre-factor, originated from difference between structure constants of group
that appear in vertex functions \cite{jab2}.

Now we come to seagull diagram, with the contribution
\bea
i{\cal M}^{\rm s.g.}_{fi}=-4ie^{2}\epsilon^{\mu}_{1} \epsilon^{\nu}_{2} \epsilon^{*
\lambda}_{3} \epsilon^{* \sigma}_{4}
&\big[&\!\!\!\!\sin\!\big(\frac{\mathbf{p}_{1}\ltimes
\mathbf{p}_{2}}{2}\big)\sin\!\big(\frac{\mathbf{p}_{3}\ltimes
\mathbf{p}_{4}}{2}\big)(g^{\mu \lambda }   g^{\nu \sigma}
-g^{\mu \sigma }g^{\nu \lambda })
\nonumber\\
&+&\!\!\sin\!\big(\frac{\mathbf{p}_{3}\ltimes \mathbf{p}_{1}}{2}\big)\sin\!\big(\frac{\mathbf{p}_{2}\ltimes \mathbf{p}_{4}}{2}\big)(g^{\mu \nu
}g^{\lambda \sigma }-g^{\mu \sigma }g^{\nu \lambda })
\nonumber\\
&+&\!\!\sin\!\big(\frac{\mathbf{p}_{1}\ltimes \mathbf{p}_{4}}{2}\big)
\sin\!\big(\frac{\mathbf{p}_{2}\ltimes \mathbf{p}_{3}}{2}\big)
(g^{\mu\nu }g^{\mu \lambda }-g^{\mu \lambda }g^{\nu \sigma})\big]
\eea
leading to
\bea
i{\cal M}^{\rm s.g.}_{fi}&=&-8ie^{2}\big[\sin\!\big(\frac{\mathbf{p}_{1}\ltimes
\mathbf{p}_{2}}{2}\big)\sin\!\big(\frac{\mathbf{p}_{3}\ltimes
\mathbf{p}_{4}}{2}\big)\epsilon_{1}  \cdot   \epsilon^{*}_{3}
\nonumber\\
&&+\sin\!\big(\frac{\mathbf{p}_{1}\ltimes \mathbf{p}_{3}}{2}\big)\sin\!\big(\frac{\mathbf{p}_{2}\ltimes \mathbf{p}_{4}}{2}\big)\times
(\epsilon_{1} \cdot
\epsilon_{2}   \epsilon^{* }_{3}\cdot \epsilon^{*}_{4}-\epsilon_{1}\cdot \epsilon^{* }_{4}\epsilon_{2} \cdot
\epsilon^{* }_{3})\big]
\eea
If we keep only zeroth orders of momentum in the bracket, we have
\bea
i{\cal M}^{\rm s.g.}_{fi}&=&-8ie^{2}\big[\sin\!\big(\frac{\mathbf{p}_{1}\ltimes
\mathbf{p}_{2}}{2}\big)\sin\!\big(\frac{\mathbf{p}_{3}\ltimes
\mathbf{p}_{4}}{2}\big)\mathbf{e}_{1}\cdot\mathbf{e}^{*}_{3}
\nonumber\\
&&+\sin\!\big(\frac{\mathbf{p}_{1}\ltimes \mathbf{p}_{3}}{2}\big)\sin\!\big(\frac{\mathbf{p}_{2}\ltimes \mathbf{p}_{4}}{2}\big)
(\mathbf{e}_{1}
\cdot \mathbf{e}_{2}\;\mathbf{e}^{*}_{3}\cdot \mathbf{e}^{*}_{4}
-\mathbf{e}_{1}\cdot \mathbf{e}^{*}_{4}\;\mathbf{e}_{2}
\cdot \mathbf{e}^{*}_{3})\big]
\eea
We can write as $\mathbf{e}_{1}\cdot \mathbf{e}_{2}\;\mathbf{e}^{*}_{3}\cdot\mathbf{e}^{*}_{4}=e^{*
l}_{3}e^{k}_{1}e^{* l}_{4}e^{k}_{2}$. Expressing $e^{*l}_{3}e^{k}_{1}=e^{*i}_{3} \delta ^{il}\delta ^{kj}
e^{j}_{1}$, and having from the spin-form $(S^{k}S^{l})_{ij}=\delta ^{kl}
\delta _{ij}-\delta ^{l}_{i} \delta^{k} _{j}$, we get
$e^{*l}_{3}e^{k}_{1}=e ^\dag _{3}(\delta^{kl}-S^{k}_{1}S^{l}_{1})e_{1}$. Using this
we obtain
\bea
\mathbf{e}_{1} \cdot \mathbf{e}_{2}\;\mathbf{e}^{*}_{3}\cdot\mathbf{e}^{*}_{4}
&=&e ^\dag _{3} e ^\dag_{4}(\delta^{kl}-S^{k}_{1}S^{l}_{1})\times
\big[(\delta^{kl}-S^{k}_{2}S^{l}_{2})-(\delta^{lk}-S^{l}_{2}S^{k}_{2})\big]e_{1}e_{2}
\nonumber\\
&=&e ^\dag _{3} e ^\dag _{4}\big[(\delta^{kl}-S^{k}_{1}S^{l}_{1})   i\epsilon ^{lkm}S ^{m}_{2}\big]e_{1}e_{2}
=e ^\dag _{3} e ^\dag
_{4}(\mathbf {S}_{1}\cdot\mathbf{S}_{2})e_{1}e_{2}
\eea
In above, we use the relation $\mathbf{S} \times \mathbf{S}=i \mathbf{S}$. Finally we have
\bea
i {\cal M}^{\rm s.g.}_{fi}=8i e^{2}
\!\sin\!\big(\frac{\mathbf{p}_{1}\ltimes \mathbf{p}_{3}}{2}\big)
\sin\!\big(\frac{\mathbf{p}_{2}\ltimes\mathbf{p}_{4}}{2}\big)
\big(\frac{1}{2}\mathbf{S}^{2}-2\big)
\eea
Now we mention that the contribution of seagull channel, for small noncommutativity parameter is something proportional to
$\big(\theta \bf{p}\big)^2 \bf{p}^2$ which is order of $\mathbf{p}^2$ that we ignore. This observation is different
from that for QCD glue-balls, for them the contribution of seagull diagram
is in zeroth order of momentum, and thus should be kept. The seagull's contribution appears to be in form of $\delta(\mathbf{r})$
in the potential (\ref{oge}).

\subsection{Effective Potential Between Photons}
Before we proceed, we define the vector $\bm{\theta}$ based on tensor $\theta^{ij}$ by
\bea
\theta ^{i}\equiv \epsilon ^{ijk}\theta _{jk}\Longrightarrow
\theta _{lm}=\frac{1}{2}\epsilon _{ilm}\theta ^{i}
\eea
By this vector we can write the $\ltimes$-product as
\bea
\mathbf{a} \ltimes \mathbf{b}=\theta^{ij}a^{i}b^{j} =a^{i}
\frac{1}{2}\;\epsilon _{lij}\theta ^{l} b^{j}
=\frac{1}{2}\;\bm{\theta}\cdot (\mathbf{a} \times \mathbf{b})
=\frac{1}{2}\;\mathbf{b}\cdot (\bm{\theta}\times \mathbf{a})
\eea
By this we have for the t-channel contribution:
\bea
i {\cal M}^{\rm t}_{fi}=4ie^2
\frac{\sin^2\!\big(\frac{1}{2}\mathbf{q}\cdot\bm{\lambda}\big)}{\mathbf{q}^{2}+m^{2}}
\Upsilon (\mathbf{q})
\eea
in which $\bm{\lambda}=\frac{1}{2}\; \mathbf{p}\times\bm{\theta}$ and
\bea
\Upsilon (\mathbf{q})=4m^{2}+3\mathbf{q}^{2}-2\mathbf{S}^{2}\mathbf{q}^{2}
+2 (\mathbf{S}\cdot\mathbf{q} )^{2}+6i\mathbf{S} \cdot  (\mathbf{q} \times
\mathbf{p} )
\eea
By the total amplitude the potential can be deduced using (\ref{potential})
\bea
V_{2\gamma}(\mathbf{r})&=&\int
\frac{{\rm d}^{3}q}{8\pi^{3}}\frac{i\e^{i\mathbf{q}\cdot\mathbf{r}}}{4\sqrt{E_{1}E_{2}E_{3}E_{4}}}\,i{\cal M}_{fi}
\nonumber\\
&=&-\frac{e^{2}}{m^{2}}\!\!\int\!
\frac{{\rm d}^{3}q}{8\pi^{3}}\frac{\e^{i\mathbf{q}\cdot\mathbf{r}}}{\mathbf{q}^{2}+m^{2}}
\sin^2\!\big(\frac{1}{2}\mathbf{q}\cdot \bm{\lambda}\big)
\Upsilon  (\mathbf{\mathbf{q}} )
\eea
By writing the exponential form of $\sin(...)$, we get
\bea
V_{2\gamma}(\mathbf{r})=-\frac{e^{2}}{4m^{2}}\int\frac{\d^{3}q}{8\pi^{3}}
\frac{\e^{i\mathbf{q}\cdot\mathbf{r}}[2-\e^{i\mathbf{q}\cdot\bm{\lambda}}
-\e^{-i\mathbf{q}\cdot\bm{\lambda}}]}{\mathbf{q}^{2}+m^{2}}
\Upsilon  (\mathbf{\mathbf{q}} )
\eea
By defining
\bea
U(R):=\int\frac{\d^{3}q}{8\pi^{3}}
\frac{\e^{i\mathbf{q}\cdot\mathbf{R}}}{\mathbf{q}^{2}+m^{2}}
=\frac{\e^{-mR}}{4\pi R}
\eea
with $R=|\mathbf{R}|$, and by $\mathbf{q}\rightarrow-i\bm{\nabla}$, we have
\bea\label{pot-0}
V_{2\gamma}(\mathbf{r})=-\frac{e^{2}}{4m^{2}}
\Upsilon(-i \bm{\nabla})
\Big[2\, U(r)-U(r_+)-U(r_-)\Big]
\eea
with $\mathbf{r}_\pm=\mathbf{r}\pm\bm{\lambda}$, and
\bea
\Upsilon(-i\bm{\nabla})=4m^{2}+(2\mathbf{S}^{2}-3)\bm{\nabla}^{2}-2(\mathbf{S}\cdot\bm{\nabla})^{2}
+6(\mathbf{p}\times\mathbf{S})\cdot\bm{\nabla}
\eea
We mention that, for $\bm{\lambda}=\mathbf{0}$ the potential vanishes; this happens in the following cases:
1) $\bm{\theta}=\mathbf{0}$, 2) $\mathbf{p}=\mathbf{0}$, and 3) $\mathbf{p}\parallel\bm{\theta}$.
It is reasonable to see the behavior of potential for small noncommutativity parameter,
defined here by $\lambda\ll r$ and $\lambda m \ll 1$. In this limit, the first surviving terms are given by:
\bea
V_{2\gamma}(\mathbf{r})=\frac{e^{2}}{4m^{2}}\Upsilon(-i \bm{\nabla})
(\bm{\lambda}\cdot\bm{\nabla})^2 U(r)
+O(\lambda^4)
\eea
Recalling that for a function $f(r)$, $\partial_i f(r)=x_i \nabla_r f$, with
$\nabla_r=r^{-1}\partial_r$, and using
\bea
(\mathbf{p}\times\mathbf{S})\cdot\mathbf{r}=(\mathbf{r}\times\mathbf{p})\cdot\mathbf{S}=\mathbf{L}\cdot\mathbf{S},\\
\nabla^2 U(r)= m^2 U(r) - \delta(\mathbf{r}),
\eea
with $\mathbf{L}$ as the total angular momentum, we get the expression for potential
\bea\label{pot-1}
V_{2\gamma}(\mathbf{r})\!\!\!&=&\!\!\!\frac{e^{2}}{4m^{2}}\Bigg\{
m^{2}\big(1+2S^2\big)\bigg[\lambda^{2}\nabla_r +\big(\bm{\lambda}\cdot\mathbf{r}\big)^2\nabla_r\nabla_r\bigg]
\nonumber\\
&&
-2\bigg[\Big[S^2\lambda^2+2\big(\bm{\lambda}\cdot\mathbf{S}\big)^{2}\Big] \nabla_r\nabla_r
+\big(\bm{\lambda}\cdot\mathbf{r}\big)^{2}\big(\mathbf{S}\cdot\mathbf{r}\big)^{2}\nabla_r\nabla_r\nabla_r\nabla_r
\nonumber\\
&&
+\Big[4\big(\bm{\lambda}\cdot \mathbf{S}\big)\big(\bm{\lambda}\cdot\mathbf{r}\big)\big(\mathbf{S}\cdot\mathbf{r}\big)
+\lambda^{2}\big(\mathbf{S}\cdot\mathbf{r}\big)^{2}+S^2\big(\bm{\lambda}\cdot \mathbf{r}\big)^2\Big]\nabla_r\nabla_r\nabla_r\bigg]
\nonumber\\
&&+6\bigg[\Big[\lambda^{2}\nabla_r\nabla_r+\big(\bm{\lambda}\cdot\mathbf{r}\big)^{2}
\nabla_r\nabla_r\nabla_r\Big]\big(\mathbf{L}\cdot\mathbf{S}\big)
\nonumber\\
&&+2\big(\mathbf{p}\times\mathbf{S}\big)\cdot\bm{\lambda}\;\big(\bm{\lambda}\cdot\mathbf{r}\big)\;\nabla_r\nabla_r\bigg]\Bigg\}\;U(r)
+{\rm D.D.} + O(\lambda^4)
\eea
in which $\lambda=|\bm{\lambda}|$, $S=|\mathbf{S}|$, and D.D. is for the distributional derivatives of function $U(r)$, containing
$\delta$-function and its derivatives; we calculate and present the explicit expression of D.D. in Appendix A.
For sake of completeness, we just present the relevant expressions
\bea
\nabla_r U&=& -\frac{\e^{ - mr}}{4\pi}\frac {mr +1}{r^{3}},\nonumber\\
\nabla_r \nabla_r U&=&\frac{\e^{ - mr}}{4\pi}\frac {m^{2}r^{2}+3mr+3}{r^{5}}\nonumber\\
\nabla_r \nabla_r \nabla_r U&=& - \frac{\e^{ - mr}}{4\pi}\frac {m^{3}r^{3}+6m^{2}r^{2}+15mr+15}{r^{7}}\nonumber\\
\nabla_r \nabla_r \nabla_r \nabla_r U &=& \frac{\e^{ - mr}}{4\pi}\frac{m^{4}r^{4}+10m^{3}r^{3}+45m^{2}r^{2}+105mr+105}{r^{9}}
\eea
We make comments on the potential given by (\ref{pot-1}). First we mention that due to $\mathbf{r}$'s in the inner products,
the effective lowest power is $r^{-5}$. Second,  the strength of potential, through the definition of
$\bm{\lambda}$, depends on momentum. Third, let us consider the spin-independent
part of the potential, that is setting $S=0$:
\bea\label{pot-2}
V_{2\gamma}^{S=0}(\mathbf{r})=\frac{e^{2}}{4}\frac{\e^{-mr}}{4\pi}
\bigg[-\lambda^{2}\frac{mr+1}{r^{3}}+(\bm{\lambda}\cdot
\mathbf{\hat{r}})^2\frac {m^{2}r^{2} +3mr+3}{r^{3}}\bigg]
\eea
We mention that $m=0$ limit of above expression is well defined. It is known that in noncommutative
field theories particles behave as electric dipoles \cite{phen-nc2, jab3, jab4,jab5, fat-moh}.
The electric dipole depends on the strength of noncommutativity parameter as well as the momentum, and is perpendicular
to both of them, and is given by $\mathbf{d}=\frac{1}{4}e\bm{\theta}\times \mathbf{p}$. For the two-photon system, in center-of-mass frame,
for which $\mathbf{p}_1=-\mathbf{p}_2=\mathbf{p}$, we have $\mathbf{d}_1=-\mathbf{d}_2=\mathbf{d}$.
The potential for a system of two electric dipoles $\mathbf{d}_1$ and $\mathbf{d}_2$ is given by
\bea\label{dipoles}
V_{\rm dipoles}(\mathbf{r})=\frac{1}{4\pi}\frac{1}{r^3}
\bigg[\mathbf{d}_1\cdot\mathbf{d}_2-3 (\mathbf{d}_1\cdot\mathbf{\hat{r}})(\mathbf{d}_2\cdot\mathbf{\hat{r}})\bigg]
\eea
We see that two expressions (\ref{pot-2}) and (\ref{dipoles}) are equivalent for $m=0$ and $\mathbf{d}=\frac{1}{2}e\bm{\lambda}$.
In fact the expression (\ref{pot-2}) is the potential of two anti-parallel dipoles in a theory in which the
potential of a charged particle is given by the so-called Yukawa potential: $V(r)=\frac{e}{4\pi}\e^{-mr}/r$.
In principle, one could justify that the potential (\ref{pot-1}) is in fact that for two anti-parallel dipoles, included by
spin-orbit and spin-dipole interactions in a Yukawa type theory. Finally, we mention:
\bea
\bm{\lambda}\cdot \mathbf{r}=\frac{1}{2}( \mathbf{p}\times\bm{\theta})\cdot\mathbf{r} =
\frac{1}{2}\bm{\theta}\cdot(\mathbf{r}\times\mathbf{p})=\frac{1}{2}\;\bm{\theta}\cdot\mathbf{L}
\eea
that can be inserted in the relevant parts of potential (\ref{pot-1}).
It is the famous $\theta$-$L$ coupling, previously found in studies concerning the implications of noncommutativity
in low energy phenomena \cite{phen-nc2,jab3,jab5}.

\section{On Existence Of Bound States}
Having the effective potential, the starting point for studying the bound state problem is the Schrodinger-type
equation by the Hamiltonian:
\bea\label{ham-1}
H=2m+H_{\rm 2b}
\eea
in which $m$ is the constituent mass, and $H_{\rm 2b}$ is sitting for the Hamiltonian capturing the dynamics of two-body system.
For example, in the glue-ball case $H_{\rm 2b}$ usually consists three parts: the kinetic term, the potential term
coming from perturbative calculation,
like (\ref{oge}), and string potential. The string potential usually is taken in the form $V_{\rm string}=2m (1-\e^{-\beta r})$, in which $\beta$ is related
to the tension of string stretched between the gluons. The formation of strings is expected from simulations on lattice, as well as the confinement
hypothesis \cite{corn-soni, hou-luo-wong,hou-wong}. Due to lack of analytical solutions, approximation methods, specially the variational method,
appear to be practically useful \cite{corn-soni, hou-luo-wong,hou-wong}. We mention that
without any reliable estimation on the value of constituent mass, all efforts for the evaluation of bound state
properties, such as mass and size, do not get any definitive result. There have been lots of
theoretical and numerical efforts, like those done using the lattice version of theory, together with
phenomenological expectations, to estimate the mass of constituent gluons.

Comparing to the case with glue-balls, the situation is more difficult in any study of photo-balls of noncommutative QED.
First, by the present experimental data we just can suggest an upper limit for noncommutative effects, leaving unspecified
$\theta$. Second, at present neither we can say anything about the value of constituent mass, nor how it varies with other
parameters, specially $\theta$. In this sense, any study can not yield definitive result or suggestion
for the quantities we like to know about photo-balls.

Here we try to formulate the dynamics based on the effective potential obtained in previous section.
Based on this formulation, we specially present a proof of existence for the bound states.
Since the issue of possible formation of string-like objects in noncommutative QED is not
in a conclusive situation, we do not consider a string potential in this work. We remind
that by including the string potential the existence proof of bound states would be a
trivial task. Also as the potential (\ref{pot-1}) is very complicated for study the possible bound states,
we restrict here ourselves to $S=0$ case; we also ignore  D.D. terms.

So we have potential (\ref{pot-2}), and for sake of definiteness, we take the vector $\bm{\theta}$
in $z$ direction, that is $\bm{\theta}=\theta\, \mathbf{z}$. It is more convenient
to work in cylindrical coordinates $(\rho,\phi,z)$, in which the kinetic energy, recalling that
the effective mass in relative motion is $m/2$, is
$T=\frac{1}{2}\frac{m}{2}(\dot{\rho}^{2}+\rho^{2}\dot{\phi}^{2}+\dot{z}^{2})$. Then we have
\bea
\lambda^{2}&=&\frac{1}{2}\epsilon^{ijk}p_{j}\theta_{k}
\frac{1}{2}\epsilon^{ilm}p_{l}\theta_{m}=\frac{1}{4}\big(\theta^2 p^2 - (\bm{\theta}\cdot\mathbf{p})^2\big)\nonumber\\
&=&\frac{1}{4}\theta^2(p_{x}^{2}+p_{y}^{2})=\frac{1}{16}m^2\theta^2(\dot{\rho}^{2}+\rho^{2}\dot{\phi}^{2})
\eea
and also
\bea
(\bm{\lambda}\cdot\hat{\mathbf{r}})^{2}=\frac{1}{r^{2}}(\bm{\lambda}\cdot\mathbf{r})^{2}
=\frac{1}{16r^2}m^2\theta^2\rho^{4}\dot{\phi}^2
\eea
in which $r$ is the distance between two photons, $r=\sqrt{\rho^{2}+z^{2}}$.
We see, while the contribution coming from the velocity $\dot{\rho}$ always yields an attractive force,
the contribution from angular velocity $\dot{\phi}$ depends on the ratio $\rho^2/r^2$, could be attractive or repulsive.
In fact the ratio $\rho^2/r^2$, as represents how much the photons move off from the plane $z=0$, also determines
the relative orientation between $\mathbf{r}$ and the components of electric dipoles generated due to
velocity $\dot{\phi}$. We recall that the relative orientation of dipoles and the position vector
appears in dipole-dipole potential (\ref{dipoles}). By these all we have the Lagrangian
\bea\label{lag-1}
L=T-V=\frac{1}{4}m\big[1+af_{1}(r)\big]\dot{\rho}^{2}+\frac{1}{4}m\dot{z}^2+
\frac{1}{4}m\rho^{2}\big[1+a\big(f_{1}(r)-\rho^{2}f_{2}(r)\big)\big]\dot{\phi}^{2}
\eea
in which $a=\frac{e^{2}}{64\pi}m\theta^{2}$ is a constant, and
\bea
f_{1}(r)=\e^{-mr}\frac{mr+1}{r^3},\;\;\;\;
f_{2}(r)=-\frac{1}{r}\frac{\partial f_1}{\partial r}=\e^{-mr}\frac{m^{2}r^{2}+3mr+3}{r^5}
\eea
We mention that the first two terms are positive definite, while third one can be negative, zero and positive.
The coordinate $\phi$ is cyclic, and hence its momentum, given by
\bea
p_{\phi}=\frac{\partial L}{\partial\dot{\phi}}=\frac{1}{2}m\rho^{2}\big[1+a\big(f_{1}(r)-\rho^{2}f_{2}(r)\big)\big]\dot{\phi}=K
\eea
is a conserved quantity, that we show with $K$; we see later that in quantum theory $K$ should be an integer.
One can find the effective theory for coordinates $\rho$ and $z$, by eliminating $\dot{\phi}$ by using the Routhian $R$ \cite{gold}, as
\bea\label{lag-2}
L_{\rho z}&=&-R=L-\dot{\phi}p_{\phi}\nonumber\\
&=&\frac{1}{4}m\big[1+af_{1}(r)\big]\dot{\rho}^{2}+\frac{1}{4}m\dot{z}^{2}-
\frac{K^{2}}{m\rho^{2}\big[1+a\big(f_{1}(r)-\rho^{2}f_{2}(r)\big)\big]}
\eea
in which we recognize the potential
\bea
V_{\rm eff}(\rho,z)=\frac{K^{2}}{m\rho^{2}\big[1+a\big(f_{1}(r)-\rho^{2}f_{2}(r)\big)\big]}
\eea
It is useful to mention the properties of $V_{\rm eff}$:
\begin{itemize}
\item It goes to $+\infty$ for $\rho=0$ and $z\neq 0$.
\item It is 0 on $\rho=z=0$.
\item It goes to $\pm\infty$ around the curve $g(\rho,z):=1+a\big(f_{1}(r)-\rho^{2}f_{2}(r)\big)=0$.
\end{itemize}
In Fig. 3 we have presented three plots of $V_{\rm eff}$ in $\rho z$-plane for $m=a=1$, $m=10a=10$, and $a=10m=10$.
We see that $V_{\rm eff}$ goes to $-\infty$ and $+\infty$ inside and outside regions defined by the curve $g(\rho,z)=0$,
respectively. We mention also, as the plots suggest, the dynamics on $z\equiv 0$ plane
is unstable; that is a small velocity $\dot z\neq 0$ hustles particles out of $z=0$ plane.

\begin{figure}[t]
\begin{center}
\includegraphics[width=0.33\columnwidth,angle=270]{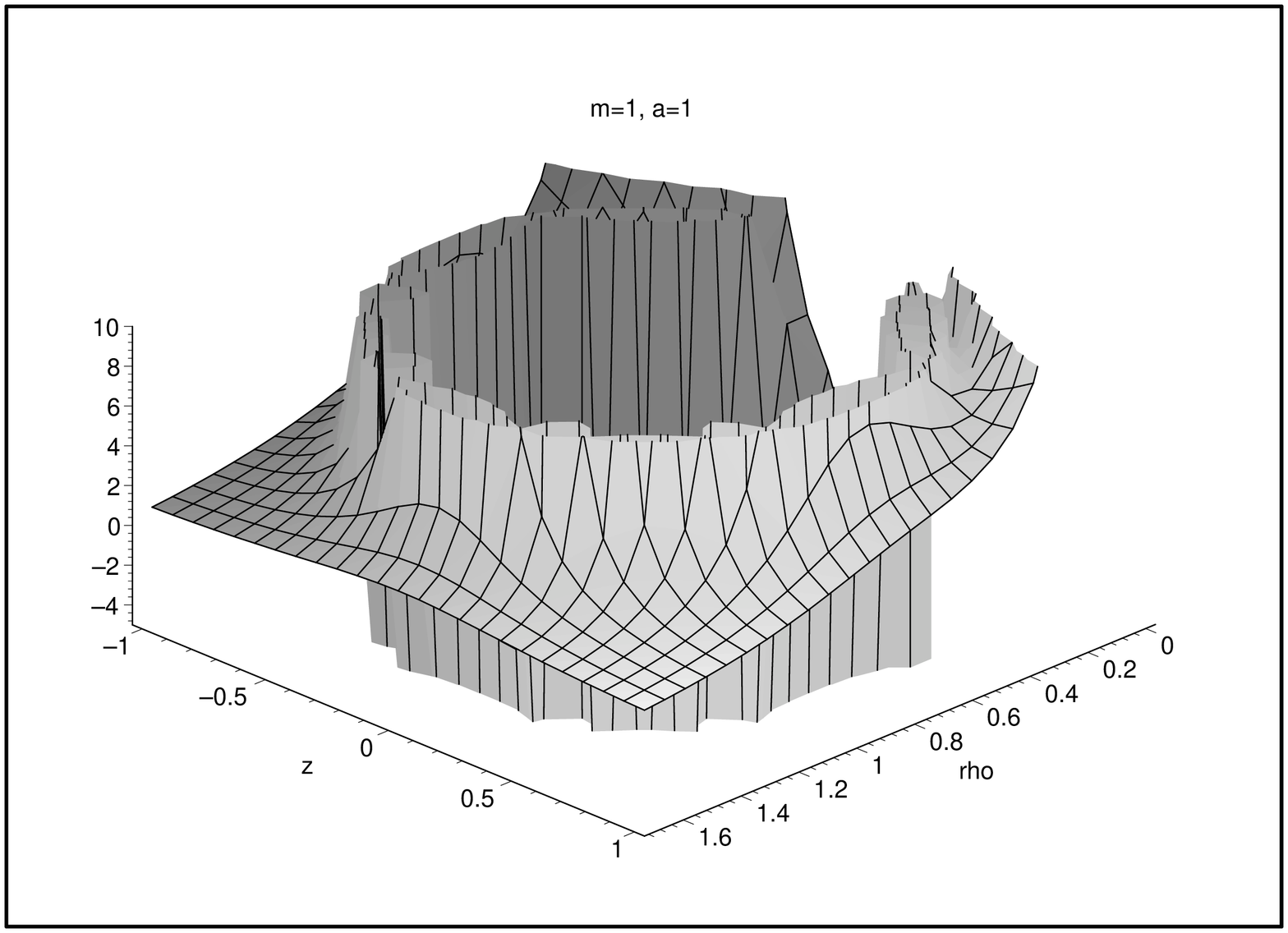}
\includegraphics[width=0.33\columnwidth,angle=270]{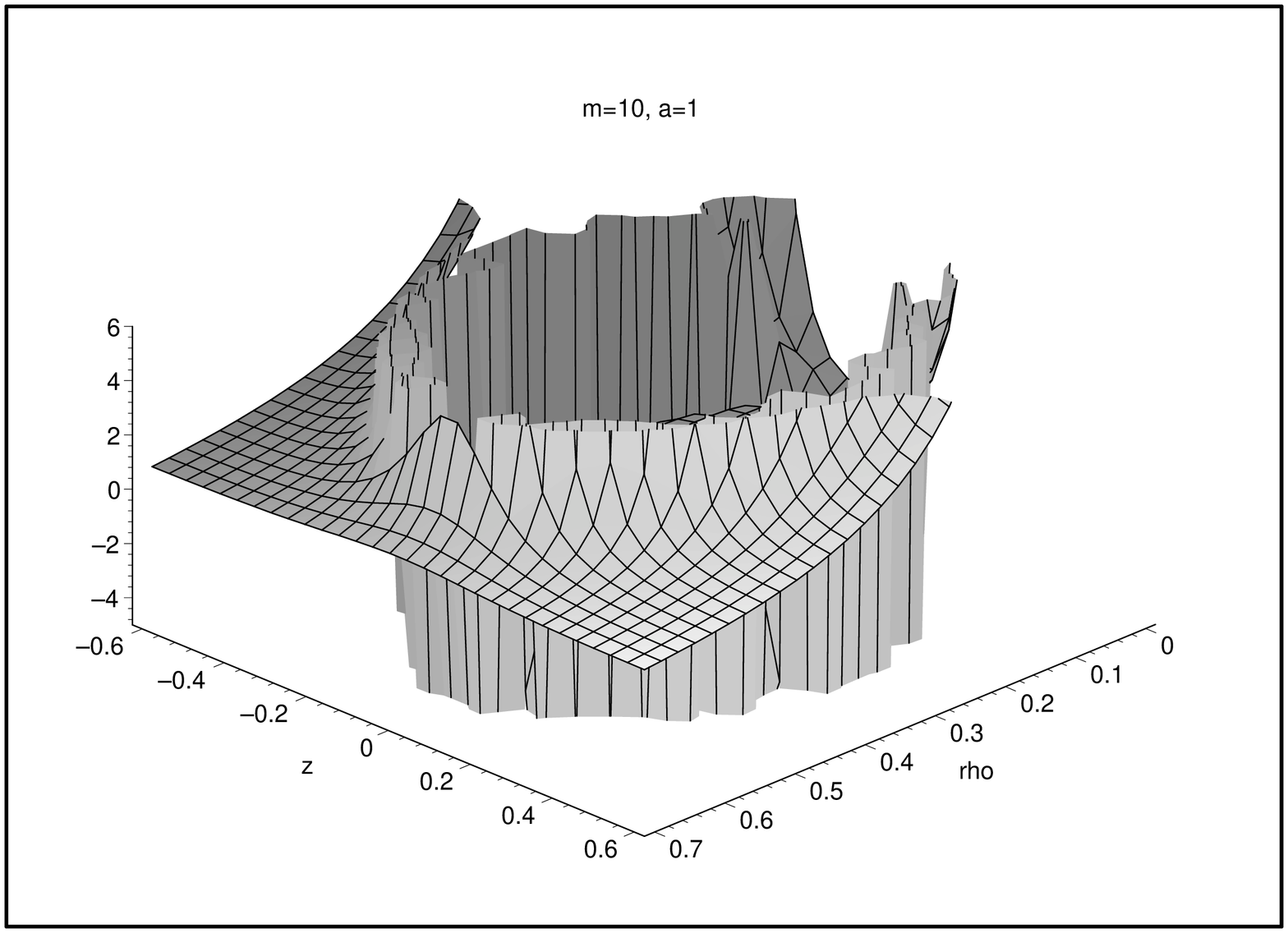}
\includegraphics[width=0.33\columnwidth,angle=270]{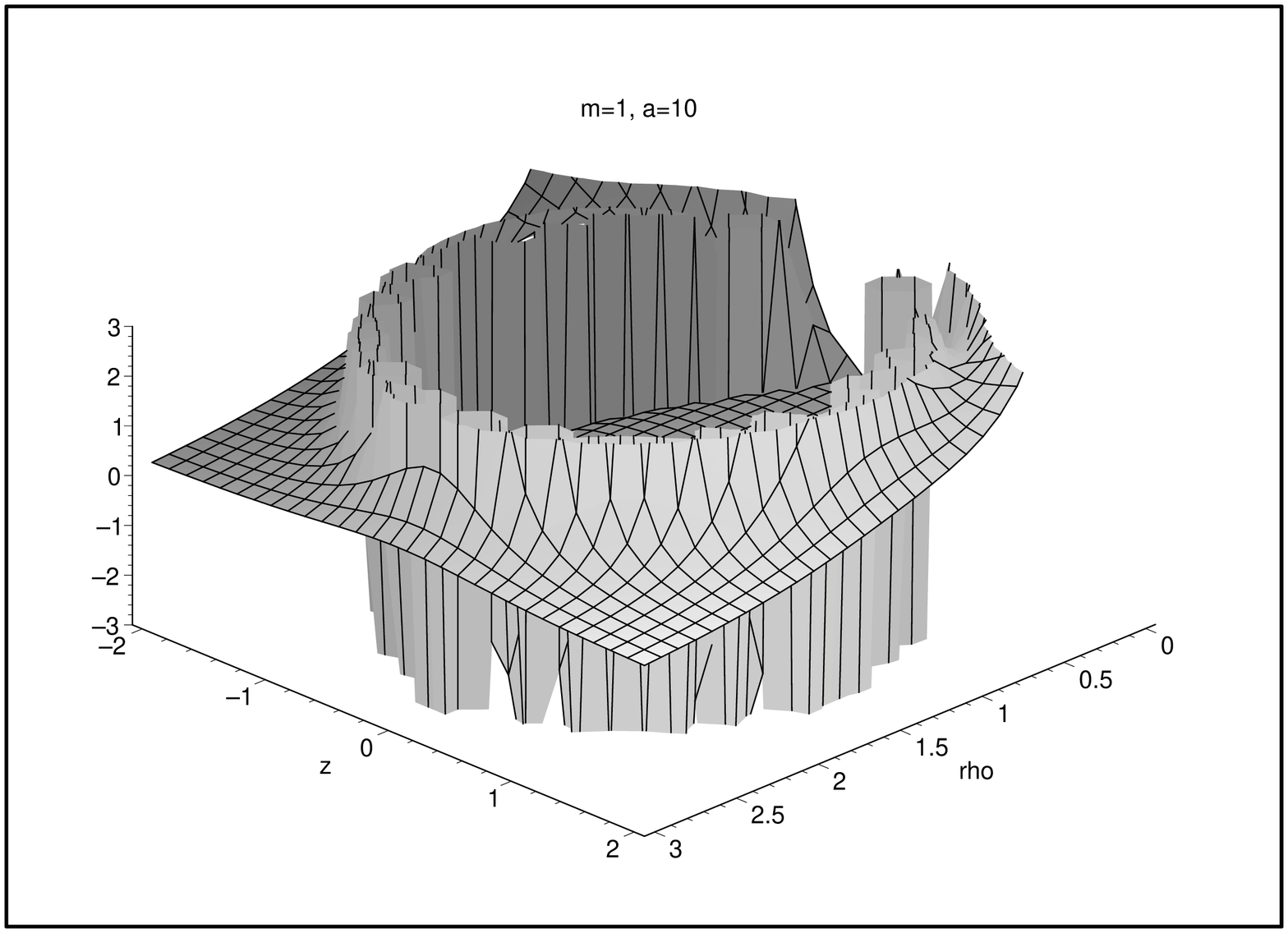}
\caption{Plots of $V_{\rm eff}$, for $m=a=1$, $m=10a=10$, and $a=10m=10$.}
\label{fig3}
\end{center}
\end{figure}

Before staring the discussion on quantum theory, let us have another look to the original Lagrangian (\ref{lag-1}).
We mention that the Lagrangian is in the form of a pure kinetic term, represented by means of a metric $g_{ij}(x)$ as
\bea
L=\frac{1}{2}\frac{m}{2} g_{ij}(x)\; \dot{x}^i \dot{x}^j,
\eea
in which $x^i=(\rho,\phi,z)$, and
\bea
g_{11}(\rho,z)=1+af_{1}(r), && \;\;\;\;g_{22}(\rho,z)=\rho^{2}\big[1+a\big(f_{1}(r)-\rho^{2}f_{2}(r)\big)\big],\nonumber\\
g_{33}=1, && \;\;\;\;\;\; g_{ij}=0,\;\; i\neq j.
\eea
We remind that although the Lagrangian is looking like a pure kinetic term, since one of the components of metric, $g_{22}$,
changes sign, we can have negative energy states, among them there are bound states. By this interpretation of Lagrangian,
the Hamiltonian of quantum theory is simply gained:
\bea
H_{\rm 2b}=-\frac{1}{m}\nabla^2=-\frac{1}{m}\frac{1}{\sqrt{|\det g|}}\partial_i\Big[\sqrt{|\det g|}\;g^{ij}\;\partial_j\Big]
\eea
in which $\deg g$ is the determinant of $g_{ij}$. As $g_{ij}$ is diagonal, $g^{ij}=1/g_{ij}$, for non-zero $g^{ij}$'s.
We mention $\deg g$ and components of $g^{ij}$ are independent of coordinate $\phi$, and so we find
\bea
H_{\rm 2b}=-\frac{1}{m}\bigg\{\frac{1}{\sqrt{g_{11}|g_{22}|}}\partial_\rho\Big[\sqrt{\frac{|g_{22}|}{g_{11}}}\;\partial_\rho\Big]
+\frac{1}{\sqrt{g_{11}|g_{22}|}}\partial_z\Big[\sqrt{g_{11}|g_{22}|}\;\partial_z\Big]+\frac{1}{g_{22}}\partial_\phi^2\bigg\}
\eea
Using the separation of variables, we choose the wave-function $\Psi(\rho,z,\phi)=\psi(\rho,z) \Phi(\phi)$, with
$\Phi(\phi)\propto\e^{il\phi}$, and due to single valued-ness of wave-function,
$l$ should be an integer. So we replace $\frac{1}{g_{22}}\partial_\phi^2$ by $-\frac{l^2}{g_{22}}$ in above, getting
\bea
H_{\rm 2b}^l=-\frac{1}{m}\bigg\{\frac{1}{\sqrt{g_{11}|g_{22}|}}\partial_\rho\Big[\sqrt{\frac{|g_{22}|}{g_{11}}}\;\partial_\rho\Big]
+\frac{1}{\sqrt{g_{11}|g_{22}|}}\partial_z\Big[\sqrt{g_{11}|g_{22}|}\;\partial_z\Big]-\frac{l^2}{g_{22}}\bigg\}
\eea
in which $H^{l}_{\rm 2b}$ means the Hamiltonian for states with specified value for $l$.
By comparison, we see that the classical counterpart of integer number $l$ is $K$. Also
we mention that $l^2/(mg_{22})$, as expected, is sitting for $V_{\rm eff}$ in quantum theory.
Now let us choose a trial-function $f(\rho,z)$, that vanishes
outside the curve $g(\rho,z)=0$. We consider the quantity:
\bea
\big\langle f\big|H^{l}_{\rm 2b}\big|f\big\rangle &=&
\int_{{\rm inside}\;g(\rho,z)=0} f^*(\rho,z)\Big(H^{l}_{\rm 2b}f(\rho,z)\Big)  \sqrt{|\det g|}\, \d\rho\, \d z \nonumber\\
&=& A_{1,f}-l^2 A_{2,f} =: E_{f,l}
\eea
in which $A_{1,f}$ and $A_{2,f}$ are two numbers independent of $l$. We mention that, since $f(\rho,z)$ vanishes in $r\to\infty$,
the contribution coming from the first two terms of $H^{l}_{\rm 2b}$, taking into account the minus sign in front, is positive.
The contribution from the last term of $H^{l}_{\rm 2b}$, reminding the definition of $f(\rho,z)$, is negative.
So for this kind of trial-function, $A_{1,f}$ and $A_{2,f}$ are positive. Here we make comment on
the existence of bound states, at least for some ranges of $l$. We mention that for sufficiently large
values of $l$, for a fixed trial-function $f(\rho,z)$, $E_{f,l}$ can be negative. In fact one can, by increasing $l$,
lower $E_{f,l}$ as much as wants. Now, by variational theorem we know that $E_{f,l}$ is an upper limit for the
lowest energy, and so we expect that for states with sufficient large $l$, there should be negative
eigenvalues for Hamiltonian $H^{l}_{\rm 2b}$. Showing these negative eigenvalues by $E_{n,l}$, and the
corresponding eigenfunctions by $\psi_{n,l}(\rho,z)$, with $n$ as for the possible quantum numbers, we have
\bea
\widetilde{\nabla}^2 \psi_{n,l}(\rho,z)=
\Big(\frac{l^2}{g_{22}}-mE_{n,l}\Big)\psi_{n,l}(\rho,z)
\eea
with $\widetilde{\nabla}^2$ as the Laplacian in $\rho z$-plane, given by
\bea
\widetilde{\nabla}^2=
\frac{1}{\sqrt{g_{11}|g_{22}|}}\partial_\rho\Big[\sqrt{\frac{|g_{22}|}{g_{11}}}\;\partial_\rho\Big]
+\frac{1}{\sqrt{g_{11}|g_{22}|}}\partial_z\Big[\sqrt{g_{11}|g_{22}|}\;\partial_z\Big]
\eea
Now, since outside of the curve $g(\rho,z)=0$ the potential $V_{\rm eff}$ is positive definite, the coefficient of $\psi_{n,l}$ in the
right-hand-side is also positive. As $r\to \infty$ belongs to the outside of the curve $g(\rho,z)=0$,
by the properties of spectrum of $\widetilde{\nabla}^2$, we expect $\psi_{n,l}|_{r\to\infty}\to0$, that is
$\psi_{n,l}$ is representing a bound state. Physically we expect that for the negative eigenvalues, the
wave-function should be localized along the well inside the curve $g(\rho,z)=0$,
as $g_{22}$ approaching zero from below.

The manner we approved the existence of bound states can be used, by increasing $l$,
for reasoning that there is no lowest energy state: the eigenvalues are unbounded from below.
We recall that the potential (\ref{pot-1}) is obtained under the assumption
that $\lambda \ll r$. As $\bm{\lambda}=\frac{1}{2}\; \mathbf{p}\times\bm{\theta}$, we
see that for large values of momentum, $\lambda$ may be comparable, and even bigger than $r$.
One situation that might invalidate the assumption $\lambda \ll r$ can happen for very large values of $l$,
corresponding to large value of $K$ in classical theory. In such cases one should consider the
original potential (\ref{pot-0}). We remind that, although the absolute least energy is meaningless
to be found under the approximation $\lambda \ll r$, the least value of energy is still meaningful
for states with specified value for $l$.

The other issue is about states with eigenvalues bigger than the maximum
of potential inside the curve $g(\rho,z)=0$. We mention that an infinite tall
wall has surrounded the inside region, and the question is if the wall can make the possibility
for forming bound states. In fact since the thickness of wall behaves like $1/h$, with $h$ as height,
by considerations coming from the WKB approximation for tunnelling effect, we expect that the particles with
positive energies can escape from the inside region. This situation is similar to the situation in one
dimensional problem with potential $V(x)=1/(x-x_0)$, for which by WKB method one finds a finite probability
expression for tunnelling of positive energy particles.

As the final point, we make comment on the possible values of spin and $l$.
The state of a two-photon system should be symmetric under the exchange of photons.
A two-photon system can have 0, 1 and 2 as total spins, as for the first and last ones the spin states are symmetric,
and for the second is anti-symmetric. Here the exchange of two photons means $z\to -z$ and $\phi\to \phi+\pi$.
By considering the spatial dependence of wave-function, we have the followings for allowed spins and $l$:
\bea
S=0,2,&~& l=0,2,4,\cdots\nonumber\\
S=1, &~& l=1,3,5,\cdots
\eea

\section{Conclusion And Discussion}
We mention that the transformations of gauge field as well as the field strength
in a noncommutative space look like to those of non-Abelian gauge theories. Besides we see that
the action of noncommutative QED contains terms which are responsible for interaction
between photons, again as the situation we have in non-Abelian gauge
theories. There is another observation that promotes the formal similarities of noncommutative and non-Abelian
theories to their behaviors, that is the negative sign of $\beta$-function, which manifests that
these theories are asymptotically free \cite{martin-ruiz, jab2}.
The above mentioned observations make it reasonable to study whether and how the photons of noncommutative QED can make
bound states. Also these observations make it reasonable to see
if the techniques developed for QCD purposes can also be used for noncommutative QED.
Here we used the so-called potential model, developed on the constituent gluon picture of QCD glue-balls.
The basic ingredient of potential model is that the self-interacting massless gauge particles may get mass
by inclusion non-perturbative effects. By calculating the amplitude for the scattering process between two massive photons, we extract
the effective potential that is expected capture the dynamics of constituent photons.
Using this effective potential, we formulate the Hamiltonian dynamics, by which arguments are presented
in favor of existence of photon bound states.

As possible photo-balls, like their glue-ball cousins, are non-perturbative in nature, it is expected that
lattice version of noncommutative QED should appear as one of the natural ways to study photo-ball's properties.
It is remarkable to remind that ordinary QED on lattice develops an area law, suggesting a stringy picture for
force, for two charged particles \cite{wilson}. There are suggestions for lattice version of noncommutative gauge
theories \cite{lat-ncqed-1}. Specially, the finite $N$ version of the theory is promising for numerical and
simulation purposes. Recently, there have been a few works reporting the preliminaries results by the lattice
version of theories \cite{lat-ncqed-2}. There are other suggestions for non-perturbative definition of noncommutative
QED \cite{stein}.

By the current experiments there has not been any signal for possible noncommutativity. So the common expectation
is that the evidence for noncommutativity, if any, should modifies the processes that occur in energies much
higher than those presently available. It is why that by present experimental data one can just suggest an
upper limit for noncommutative effects. There has been another suggestion that the noncommutativity effects may
appear due to applying sufficiently strong magnetic field on samples containing moving charged particles.
It would be extremely interesting if noncommutative view let us know something new about relevant phenomena \cite{jackiw}.

\vspace{0.5cm}
{\bf Acknowledgement:}
A. H. F. is grateful to M. Khorrami for very helpful discussions on the distributional derivatives,
and also for extremely useful discussions on the bound state problem.

\appendix

\section{Distributional Derivatives}
Here we calculate the distributional derivatives.
One very helpful reference is \cite{stackgold}. First we consider
\bea
\partial_i\partial_j\frac{\e^{-mr}}{r}
\eea
The distributional derivative can be calculated by its effect on a test-function $\phi(\mathbf{r})$
\bea
\Big\langle\partial_i\partial_j\frac{\e^{-mr}}{r},\phi\Big\rangle&:=&
\Big\langle\frac{\e^{-mr}}{r},\partial_i\partial_j\phi\Big\rangle=\!\int\frac{\e^{-mr}}{r}\;\partial_i\partial_j\phi(\mathbf{r})\dr\nonumber\\
&=&\!\!\!\!\lim_{\varepsilon\to0^+}\int_{r\geq\varepsilon}\!\!\!\frac{\e^{-mr}}{r}\;\partial_i\partial_j\phi(\mathbf{r})
\dr
\eea
in which $\dr=r^2\d r \d \Omega$. The limit above does exist because the integral
in the second line, due to $r^2$ in $\dr$, is finite. $r=0$ is excluded from the last integral, and
so we can use the identity
\bea
\frac{\e^{-mr}}{r}\partial_i\partial_j\phi=
\partial_i\Big(\frac{\e^{-mr}}{r}\partial_j\phi\Big)
-\partial_i\frac{\e^{-mr}}{r}\;\;\partial_j\phi
\eea
by which we have for $I_{ij}=\Big\langle\partial_i\partial_j\frac{\e^{-mr}}{r},\phi\Big\rangle$
\bea
I_{ij}&=&\lim_{\varepsilon\to 0^+}\bigg[\int_{r\geq\varepsilon}\partial_i\Big(\frac{\e^{-mr}}{r}\partial_j\phi\Big)\dr\!
-\int_{r\geq\varepsilon} \partial_i\frac{\e^{-mr}}{r}\;\partial_j\phi\, \dr \bigg]\nonumber\\
&=&\lim_{\varepsilon\to 0^+}\bigg[\int_{r\geq\varepsilon}
\bm{\nabla}\cdot\Big(\ehat_i\frac{\e^{-mr}}{r}\partial_j\phi\Big)\dr\!
-\int_{r\geq\varepsilon}  \partial_i\frac{\e^{-mr}}{r}\partial_j\phi\, \dr \bigg]\nonumber\\
&=&\lim_{\varepsilon\to 0^+}\bigg[\!\int_{r=\varepsilon}
\frac{\e^{-mr}}{r}\partial_j\phi (-\ehat_i\cdot \rhat)r^2\d\Omega
-\int_{r\geq\varepsilon} \partial_i\frac{\e^{-mr}}{r}\partial_j\phi\, \dr  \bigg]\nonumber\\
&=&\lim_{\varepsilon\to 0^+}\bigg[-\int_{r=\varepsilon}
\frac{\e^{-mr}}{r}\partial_j\phi\; n_ir^2\d\Omega
-\int_{r\geq\varepsilon}\partial_i\frac{\e^{-mr}}{r}\partial_j\phi\, \dr  \bigg]
\eea
in which $\ehat_i$ is for unit vector in $x$, $y$ and $z$ directions, and $\rhat=(n_1,n_2,n_3)$. The first integral
in last line is proportional to $\varepsilon$ and so vanishes in the limit. So we get
\bea
I_{ij}=-\lim_{\varepsilon\to 0^+} \int_{r\geq\varepsilon}\! \partial_i\frac{\e^{-mr}}{r}\;\partial_j\phi \;\dr
\eea
By repeating the steps above, we arrive at:
\bea
I_{ij}&=&\lim_{\varepsilon\to 0^+}\bigg[\int_{r=\varepsilon}
\phi\,\partial_i\frac{\e^{-mr}}{r}\; n_j r^2\d\Omega
+\int_{r\geq\varepsilon}\phi\,\partial_i\partial_j\frac{\e^{-mr}}{r}\, \dr  \bigg]\nonumber\\
&=&\lim_{\varepsilon\to 0^+}\bigg[-\phi(\varepsilon)\int n_in_j \d\Omega
+\int_{r\geq\varepsilon} \phi\,\partial_i\partial_j\frac{\e^{-mr}}{r}\, \dr  \bigg]
\eea
for getting it we used $\partial_i f(r)= n_i \partial_r f(r)$, and also keeping the integrand
of first integral to the first non-vanishing order in $\varepsilon$. We know
\bea
\int n_in_j\d \Omega=A_2 \delta_{ij}
\eea
and $A_2$ can be calculated by taking trace of the both sides, yielding $A_2=\frac{4\pi}{3}$.
By these all we have
\bea
I_{ij}=-\frac{4\pi}{3} \phi(0)\delta_{ij} +\lim_{\varepsilon\to 0^+}\!\int_{r\geq\varepsilon}
\!\!\!\phi\;\partial_i\partial_j\frac{\e^{-mr}}{r}\, \dr
\eea
The limit above exists, by using the fact that the value of function in origin is constant and independent
from the $\Omega=\Omega(\theta,\varphi)$, and recalling
\bea
\int (3n_in_j -\delta_{ij})\d\Omega=0
\eea
One can remove the limit by respecting the order of integrations. By these all we get
\bea
\partial_i\partial_j\frac{\e^{-mr}}{r}\to -\frac{4\pi}{3} \delta_{ij}\delta(\mathbf{r})
+{\rm pf}\Big[\partial_i\partial_j\frac{\e^{-mr}}{r}\Big]
\eea
in which ``~pf~" is sitting for pseudo-function, that here simply means that in integrals
the integration on solid-angle should be done before radial one.

Now we come to
\bea
I_{ijk}&=&\Big\langle\partial_i\partial_j\partial_k\frac{\e^{-mr}}{r},\phi\Big\rangle:=
-\Big\langle\frac{\e^{-mr}}{r},\partial_i\partial_j\partial_k\phi\Big\rangle\nonumber\\
&=&\!\!-\int\frac{\e^{-mr}}{r}\;\partial_i\partial_j\partial_k\phi(\mathbf{r})\dr\nonumber\\
&=&\!\!-\lim_{\varepsilon\to0^+}\int_{r\geq\varepsilon}\!\!\!\frac{\e^{-mr}}{r}\;\partial_i\partial_j\partial_k\phi(\mathbf{r})\dr
\eea
in which again the limit exists because the integral
in the second line is finite. By repeating the steps for calculation $I_{ij}$,
and the replacement
\bea
\phi(\varepsilon)=\phi(0)+\varepsilon \,\rhat\cdot\bm{\nabla}\phi(0)+O(\varepsilon^2)
\eea
and using
\bea
&&\int n_i n_j n_k n_p \d\Omega=A_4 \big[\delta_{ij}\delta_{kp}+\delta_{ik}\delta_{jp}+\delta_{ip}\delta_{jk} \big]\nonumber\\
&&\int n_i\d\Omega=\int n_i n_j n_k \d\Omega=0,
\eea
with $A_4=\frac{4\pi}{15}$, one reaches to
\bea
I_{ijk}&=&\frac{4\pi}{5}
\Big[ \delta_{ij}\partial_k\phi(0)+\delta_{jk}\partial_i\phi(0)+\delta_{ki}\partial_j\phi(0)\Big]
+\lim_{\varepsilon\to 0^+}\!\int_{r\geq\varepsilon}
\!\!\!\phi\;\partial_i\partial_j\partial_k\frac{\e^{-mr}}{r}\, \dr
\eea
for which again the limit above exists while one is careful that the integration on $\d\Omega$ should be done firstly.
By these all we get
\bea
\partial_i\partial_j\partial_k\frac{\e^{-mr}}{r}&\to& -\frac{4\pi}{5}
\Big[ \delta_{ij}\partial_k+\delta_{jk}\partial_i+\delta_{ki}\partial_j\Big]\delta(\mathbf{r})
+\;{\rm pf}\Big[\partial_i\partial_j\partial_k\frac{\e^{-mr}}{r}\Big]
\eea
in which again ``~pf~" simply means in integrals the integration on solid-angle should be done before radial one.

Now we come to
\bea
I_{ijkl}&=&\Big\langle\partial_i\partial_j\partial_k\partial_l\frac{\e^{-mr}}{r},\phi\Big\rangle:=
\Big\langle\frac{\e^{-mr}}{r},\partial_i\partial_j\partial_k\partial_l\phi\Big\rangle\nonumber\\
&=&\!\!\int\frac{\e^{-mr}}{r}\;\partial_i\partial_j\partial_k\partial_l\phi(\mathbf{r})\dr\nonumber\\
&=&\!\!\lim_{\varepsilon\to0^+}\int_{r\geq\varepsilon}\!\!\!\frac{\e^{-mr}}{r}\;\partial_i\partial_j\partial_k\partial_l\phi(\mathbf{r})\dr
\eea
in which again the limit exists because the integral
in the second line is finite. By repeating the steps for calculation $I_{ij}$ and $I_{ijk}$,
and the replacement
\bea
\phi(\varepsilon)=\phi(0)+\varepsilon \,\rhat\cdot\bm{\nabla}\phi(0)
+\frac{\varepsilon^2}{2} (\rhat\cdot\bm{\nabla})^2\phi(0)+O(\varepsilon^3)
\eea
and using
\bea
\int n_i n_j n_k n_p n_q \d\Omega&=&0\nonumber\\
\int n_i n_j n_k n_l n_p n_q \d\Omega&=&A_6\big[\delta_{ij}\delta_{kl}\delta_{pq}+\delta_{ik}\delta_{jl}\delta_{pq}+\delta_{il}\delta_{kj}\delta_{pq}\nonumber\\
     &&\quad+\delta_{ij}\delta_{kp}\delta_{lq}+\delta_{ij}\delta_{kq}\delta_{lp}+\delta_{ik}\delta_{jp}\delta_{lq}\nonumber\\
     &&\quad+\delta_{ik}\delta_{jq}\delta_{lp}+\delta_{il}\delta_{kp}\delta_{jq}+\delta_{il}\delta_{kq}\delta_{jp}\nonumber\\
     &&\quad+\delta_{ip}\delta_{kl}\delta_{jq}+\delta_{ip}\delta_{kj}\delta_{lq}+\delta_{ip}\delta_{kq}\delta_{jl}\nonumber\\
     &&\quad+\delta_{iq}\delta_{kl}\delta_{jp}+\delta_{iq}\delta_{kj}\delta_{lp}+\delta_{iq}\delta_{kp}\delta_{jl}\big]
\eea
with $A_6=\frac{4\pi}{17\times 9}$, one reaches to
\bea
I_{ijkl}&=&-\frac{4\pi}{15}
\Big[ \delta_{ij}\delta_{kl}+\delta_{ik}\delta_{jl}+\delta_{il}\delta_{jk}\Big]\phi(0)
-\frac{10\pi}{51}\Big[ \delta_{ij}\delta_{kl}+\delta_{ik}\delta_{jl}+\delta_{il}\delta_{jk}\Big]\nabla^2\phi(0)\nonumber\\
&&-\frac{20\pi}{51}\Big[ \delta_{ij}\partial_k\partial_l+\delta_{ik}\partial_j\partial_l+\delta_{il}\partial_j\partial_k
+\delta_{kl}\partial_i\partial_j+\delta_{jl}\partial_i\partial_k+\delta_{kj}\partial_i\partial_l\Big]\phi(0)\nonumber\\
&&+\lim_{\varepsilon\to 0^+}\!\int_{r\geq\varepsilon}
\!\!\!\phi\;\partial_i\partial_j\partial_k\partial_l\frac{\e^{-mr}}{r}\, \dr
\eea
for which again the limit above exists while one is careful that the integration on $\d\Omega$ should be done firstly.
By these all we get
\bea
\partial_i\partial_j\partial_k\partial_l\frac{\e^{-mr}}{r}&\to&
-\frac{4\pi}{15} \Big[ \delta_{ij}\delta_{kl}+\delta_{ik}\delta_{jl}+\delta_{il}\delta_{jk}\Big]\delta(\mathbf{r})
-\frac{10\pi}{51}\Big[ \delta_{ij}\delta_{kl}+\delta_{ik}\delta_{jl}+\delta_{il}\delta_{jk}\Big]\nabla^2\delta(\mathbf{r})\nonumber\\
&&-\frac{20\pi}{51}\Big[ \delta_{ij}\partial_k\partial_l+\delta_{ik}\partial_j\partial_l+\delta_{il}\partial_j\partial_k
+\delta_{kl}\partial_i\partial_j+\delta_{jl}\partial_i\partial_k+\delta_{kj}\partial_i\partial_l\Big]\delta(\mathbf{r})\nonumber\\
&&+{\rm pf}\Big[\partial_i\partial_j\partial_k\partial_l\frac{\e^{-mr}}{r}\Big]
\eea
in which again ``~pf~" simply means in integrals the integration on solid-angle should be done before radial one.
The combination $\nabla^2\partial_i\partial_j$ is simply $\delta^{kl}\partial_i\partial_j\partial_k\partial_l$.


\end{document}